# Large-Signal Stability Criteria in DC Power Grids with Distributed-Controlled Converters and Constant Power Loads

Fangyuan Chang[1], *Student Member, IEEE*, Xiaofan Cui[2], *Student Member, IEEE*
Mengqi Wang[3], *Senior Member, IEEE*, Wencong Su[4], *Senior Member, IEEE*, Alex Q. Huang[5], *Fellow, IEEE*

***Abstract*—The increasing adoption of power electronic devices may lead to large disturbance and destabilization of future power systems. However, stability criteria are still an unsolved puzzle, since traditional small-signal stability analysis is not applicable to power electronics-enabled power systems when a large disturbance occurs, such as a fault, a pulse power load, or load switching. To address this issue, this paper presents for the first time the rigorous derivation of the sufficient criteria for large-signal stability in DC microgrids with distributed-controlled DC-DC power converters. A novel type of closed-loop converter controllers is designed and considered. Moreover, this paper is the first to prove that the well-known and frequently cited Brayton-Moser's mixed potential theory (published in 1964) is incomplete. Case studies are carried out to illustrate the defects of Brayton-Moser's mixed potential theory and verify the effectiveness of the proposed novel stability criteria.**

## I. Introduction

POWER systems are going through a paradigm shift from electric machine-based to power electronics-based, with a huge number of different players on the supply side [1]-[3]. Nowadays, thousands of distributed energy resources (DERs) are being integrated into power systems through power electronics components such as solar panels, wind turbines, and energy storage systems; however, the integration of numerous power electronic components and constant power loads (CPLs) destabilizes power systems and leads to critical oscillations. Consequently, one of the crucial challenges of this new paradigm is to keep the whole power system stable. The stability issues faced by DC microgrids are especially severe and urgent due to their unique properties. First, the low inertia of DC microgrids sharply weakens their stability; and second, owing to their advantage of smooth control, DC microgrids are unprecedentedly more promising than AC power systems given the increasing penetration of DERs. Therefore, the purpose of this paper is to solve the stability issues in power-converter-dominated DC microgrids.

Recent works related to stability analysis in DC microgrids can be categorized according to the type of disturbance and the number of converters, as shown in Table I. Most of the stability studies of DC microgrids are performed using *small-signal and linearized models*, especially for large-scale DC microgrids with multiple converters and CPLs. However, linearized models of microgrids are not always applicable. The first reason is that from the perspective of a dynamic system, the power converter dynamics can be approximated by a nonlinear state-space averaging model only if the system bandwidth is well below the switching frequency [4]. The challenge here is that the feasible region of the averaging model shrinks sharply when we perform linearization for nonlinear systems with high bandwidth. Moreover, when nonlinear controllers are applied in power converters, the system dynamics become even more complicated. The second reason is that even though the small-signal approach is proven to be effective in some cases, it does not work well when a large disturbance occurs. The small-signal-based approach often utilizes classical eigenvalues or impedance techniques [5][6], with linearization of nonlinear systems and analysis of equilibrium points. The work in [7] explores small-signal stability issues in a simplified cascade distributed power architecture with a one-line regulating converter using phase portraits. Paper [8] analyzes the factors that cause the instability of a DC microgrid with multiple converters and presents two stabilization methods. In paper [9], a converter-based DC microgrid is studied by employing a multistage configuration. The authors derive a comprehensive small-signal model to analyze the interface power converters in each stage and propose virtual impedance-based stabilizers to enhance the damping of DC microgrids.

TABLE I
Classification of Recent Works about Stability Analysis in Grids

| Classification | Type of disturbance | Small-signal | Large-signal |
|---|---|---|---|
| Number of converters considered | Zero/Single converter | [7][10][13] | [12][13][14] |
| | Multiple converters | [8][9][11] | N/A |

*Large-signal stability* criteria determine the safe operation regions of real power systems. A practical application of the stability criteria is to ensure safe operation in the event of a large disturbance, which is possible in the real operation of DC microgrids, such as load switching, pulse power loads, and faults. A large-signal stable system is naturally small-signal stable; however, the opposite holds only when special prerequisites are satisfied. Some studies covering large-signal stability in recent years are discussed as follows. However, some large-signal analysis tools introduced in the literature either have limited applicable ranges or non-rigorous theoretical foundations. In [12], large-signal stability is studied in an electrical system with a single converter based on Takagi-Sugeno multi-modeling [15]. Paper [13] presents the destabilizing effect of CPLs on DC microgrids and analyzes both their small-signal stability and large-signal stability, showing a significant difference between them. That said, only one single source and CPL are considered. The work in [14] focuses on the large disturbance scenarios in a cascaded system, which represents the basic form of a DC microgrid. The authors analyze the stability of the cascaded system based on Brayton-Moser's mixed potential theory [16] and develop it under the consideration of conservatism caused by transient response characteristics of the load converter. Paper [17] presents large-signal stability criteria based on Brayton–Moser's mixed potential of a DC electrical system with multistage *LC* filters and a CPL. However, the conclusions in [14][17] may not be sound; our paper verifies that their deployment of Brayton-Moser's mixed potential theory actually cannot obtain sufficient criteria for nonlinear circuit networks. Moreover, we believe that the authors in [18] do not accurately understand Brayton-Moser's mixed potential theory when they apply it to deal with large-signal stability issues. Their definition of potential is questionable due to its violation of the basic property of potential—that is, potential depends only on the start point and endpoint, independent of the state trajectories.

[1] F. Chang is with the University of Michigan-Dearborn, MI 48128 USA (e-mail: fychang@umich.edu)
[2] X. Cui is with the University of Michigan-Ann Arbor, MI 48109 USA (e-mail: cuixf@umich.edu)
[3] M. Wang is with the University of Michigan-Dearborn, MI 48128 USA (e-mail: mengqiw@umich.edu)
[4] W. Su is with the University of Michigan-Dearborn, MI 48128 USA (e-mail: wencong@umich.edu)
[5] A.Q. Huang is with The University of Texas at Austin, TX 78712 USA (e-mail: aqhuang@utexas.edu)
Corresponding Author: Wencong Su (e-mail: wencong@umich.edu)

In a nutshell, large-signal stability criteria for DC microgrids with multiple converters are still an unsolved puzzle. For the first time, this paper presents a systematic and rigorous methodology to deal with this problem. The main contributions of this paper can be summarized as follows:

1) To the best of the authors' knowledge, we are the first to present the rigorous derivation of the sufficient criteria for large-signal stability in DC microgrids with multiple power converters and CPLs. It is worth mentioning that this derivation works for many different types of power converters.

In our DC microgrids model, the novel proposed distributed *closed-loop converter controllers* are considered. It refers to the feedback control between converter parameters (e.g., the equivalent impedance of converter) and the operation parameters of DC microgrids (e.g., node voltage).

In the real operation of DC microgrids, in order to smooth power flow and provide electric power of higher quality, it is common to regulate output voltages through the control of power converters. Therefore, it is necessary to acquire the stability criteria in DC microgrids with controlled power converters, which can be treated as a rule of thumb for the stable operation of modern DC microgrids.

2) A novel current-mode control method is proposed to regulate node voltages in DC microgrids. It shows superior performance over that of droop control in terms of stability and steady-state error.

3) We discuss and debunk the defects of the well-known Brayton-Moser's mixed potential theory [16] and conclude that it may not obtain the sufficient criteria of nonlinear circuit networks. The findings reveal several flawed studies based on this theory since the theory was proposed in 1964.

4) We investigate the superiority of the proposed large-signal stability region over the traditional small-signal stability region in DC microgrids. It is observed that the small-signal stability region of DC microgrids with high nonlinearity is not reliable in our case study.

The structure of this paper is organized as follows: In section II, the model of a typical DC microgrid with multiple power converters and CPLs is discussed. In section III, we propose a novel current-mode converter controller in DC microgrids. Section IV presents the sufficient conditions for large-signal stability in DC microgrids with distributed-controlled converters. In Section V, we reveal the defects of Brayton-Moser's potential theory and verify the correctness of our methodology. Besides, we compare the large-signal stability region solved by the novel proposed methodology and the traditional small-signal stability region. The conclusion and future work are indicated in section VI.

## II. Model Assumptions & Problem Description

The circuit structure of a generalized DC microgrid with multiple converters and CPLs is described in Fig. 1.

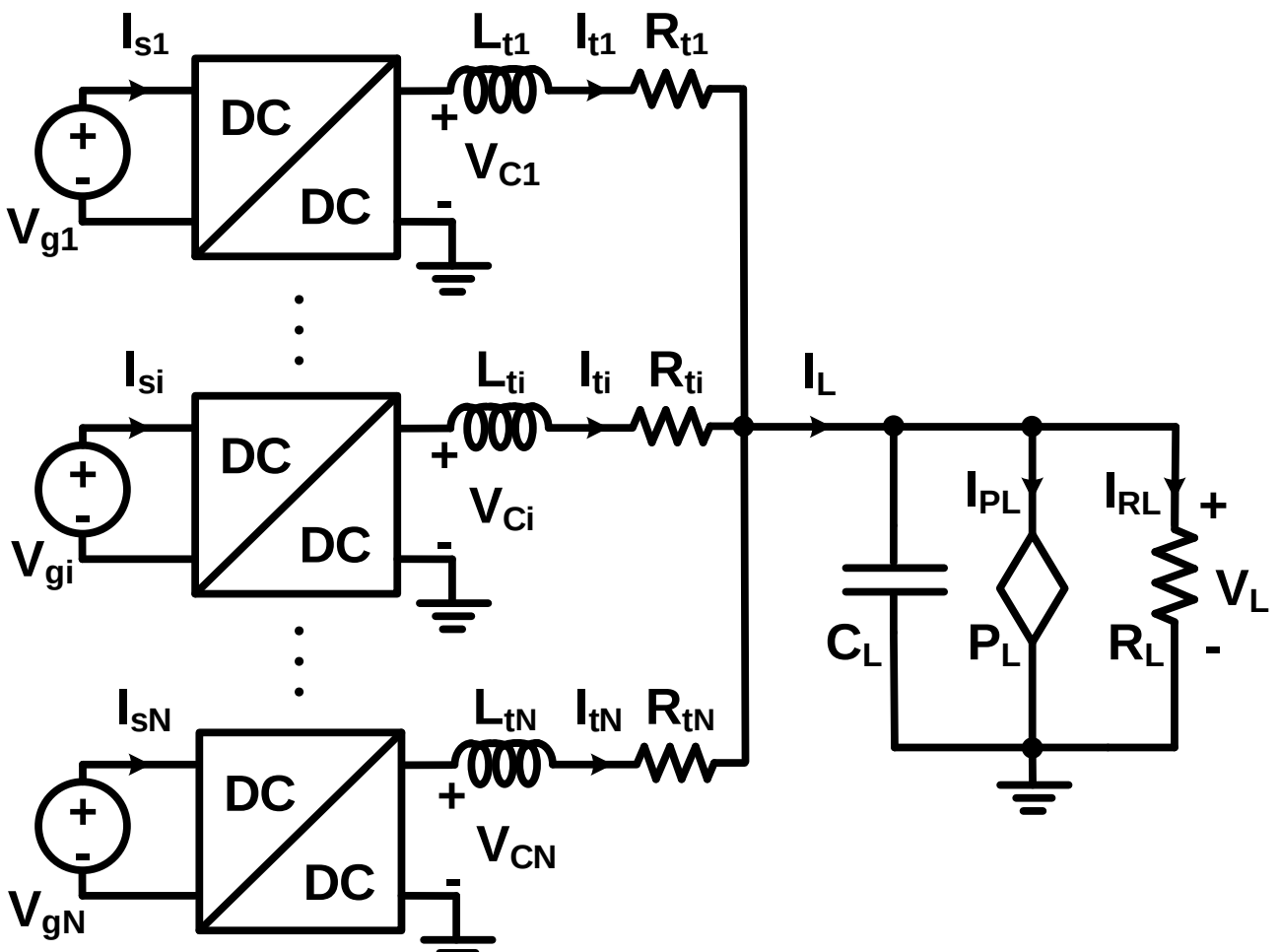

Fig. 1. The circuit structure of a typical DC microgrid with a CPL.

Without loss of generality, the circuit structure is modeled based on the following assumptions:

1) The power supplies are all constant voltage sources.
2) The DC-DC converters are employed to step up/down the voltage outputs. They can be ideal buck converters or boost converters. No parasitic resistance or parasitic capacitance is considered.
3) Every transmission line is modeled as impedance.
4) The demand side consists of an aggregated CPL and a linear resistor. The operation function of the CPL is described as the following equation, which is also depicted in Fig. 2.

$$\begin{cases} I_{PL} = I_{max}, & V_L \le V_{min} \\ V_L = P_L / I_{PL}, & V_{min} \le V_L \le V_{max} \\ V_L = V_{max}, & I_{PL} < I_{min} \end{cases} \tag{1}$$

where $I_{PL}$ and $V_L$ are the current and output voltage of the CPL, separately. $P_L$ is the power of the CPL when $V_{min} \leq V_L \leq V_{max}$. $V_{min}$ and $V_{max}$ are the lower bound and upper bound of output voltage, separately. $I_{min}$ and $I_{max}$ are the lower bound and upper bound of current, separately.

A large disturbance often happens when a fault, a pulse power load, or load switching occurs in a DC microgrid. Unfortunately, traditional small-signal stability analysis cannot provide sufficient information to determine the stability of a microgrid after such a large disturbance. In this paper, novel large-signal stability criteria are proposed to solve this issue. *Large-signal stability* is defined based on the definition of Lyapunov global asymptotic stability: There exists at least one stable equilibrium point of the dynamic system where any subsequent trajectories of the set of initial conditions end up. It guarantees that a DC microgrid will always be stable even after going through a severe disturbance.

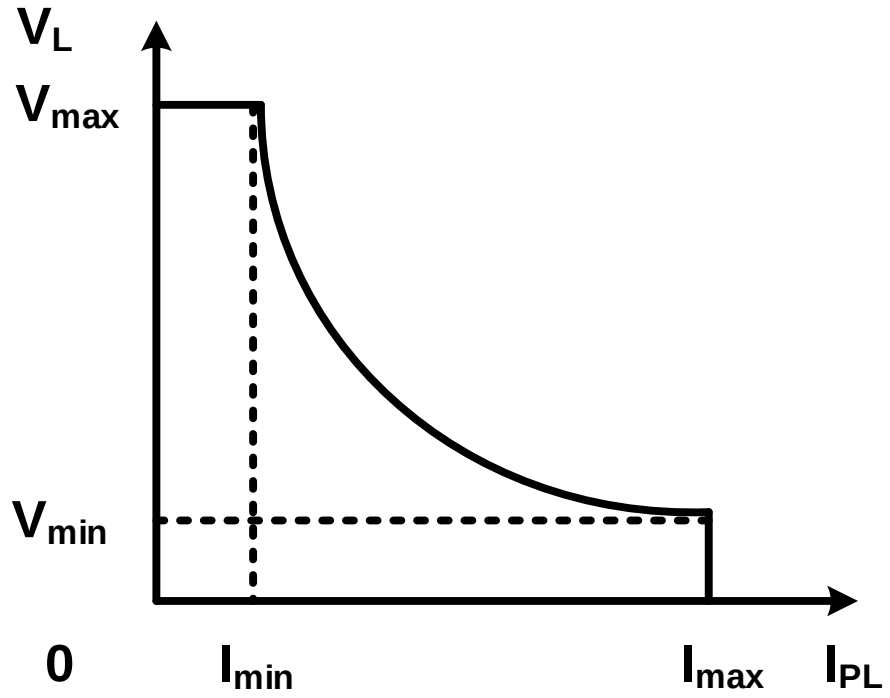

Fig. 2. The CPL operation model.

## III. The Modeling of DC Microgrids with Closed-Loop Converter Controllers

In DC microgrids, reasonable control of power converters enables the regulation of output voltages to smooth the power flow and provide electric power with high quality. Recently, different schemes of current-mode control for converters have been studied due to its unique advantages in current regulation [4][19]. Here we suppose that the power converters in the microgrid are distributed-controlled in current mode. Considering the characteristics of the output port of the switch network of the power converters, regardless of whether they are buck converters, boost converters, or buck-boost converters, the microgrid can be modeled as in Fig. 3, where $I_{si}(i = 1,2, \dots, N)$ represents a current source. A detailed explanation of the switch modeling of power converters can be found in [20].

The purpose of the distributed control of power converters is to regulate capacitor voltage $V_{Ci}$ to an expected value $V_{refi}$, through the switching of $I_{si}$ in each branch. Traditionally, droop controllers are often utilized to achieve this purpose. Here, we propose a novel type of feedback controller and deploy it in DC microgrids instead of traditional droop controllers. The model of a DC microgrid with the proposed converter controllers is depicted in Fig. 4. In each power converter, the transfer function of the controller block shown in Fig. 5 is specified as follows:

$$G(s) = \frac{I_{si}(s)}{V_{refi}(s) - V_{Ci}(s)} = Y_{in}(s) = \frac{R_{pi} + sL_{qi} + R_{qi}}{R_{pi}\left(sL_{qi} + R_{qi}\right)} = \frac{1}{R_{pi}} \frac{s + \frac{R_{pi} + R_{qi}}{L_{qi}}}{s + \frac{R_{qi}}{L_{qi}}} \quad (2)$$

where $Y_{in}$ is the equivalent admittance of the block shown in Fig. 5. Usually, $R_{qi}$ is designed to be a small resistor, and $R_{pi}$ is designed to be larger than $R_{qi}$, i.e., $R_{pi} \gg R_{qi}$.

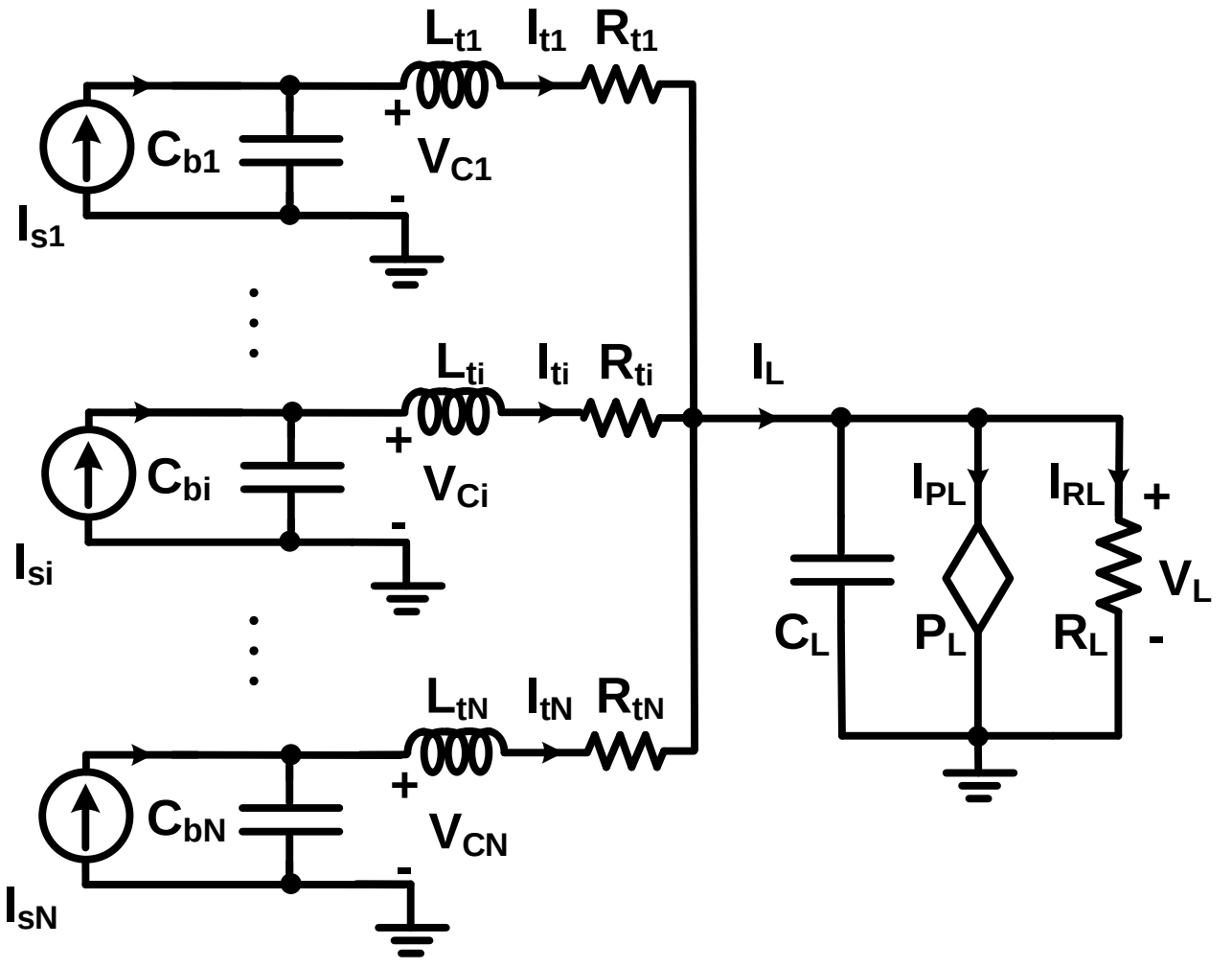

Fig. 3. The diagram of a typical DC microgrid under current-mode control.

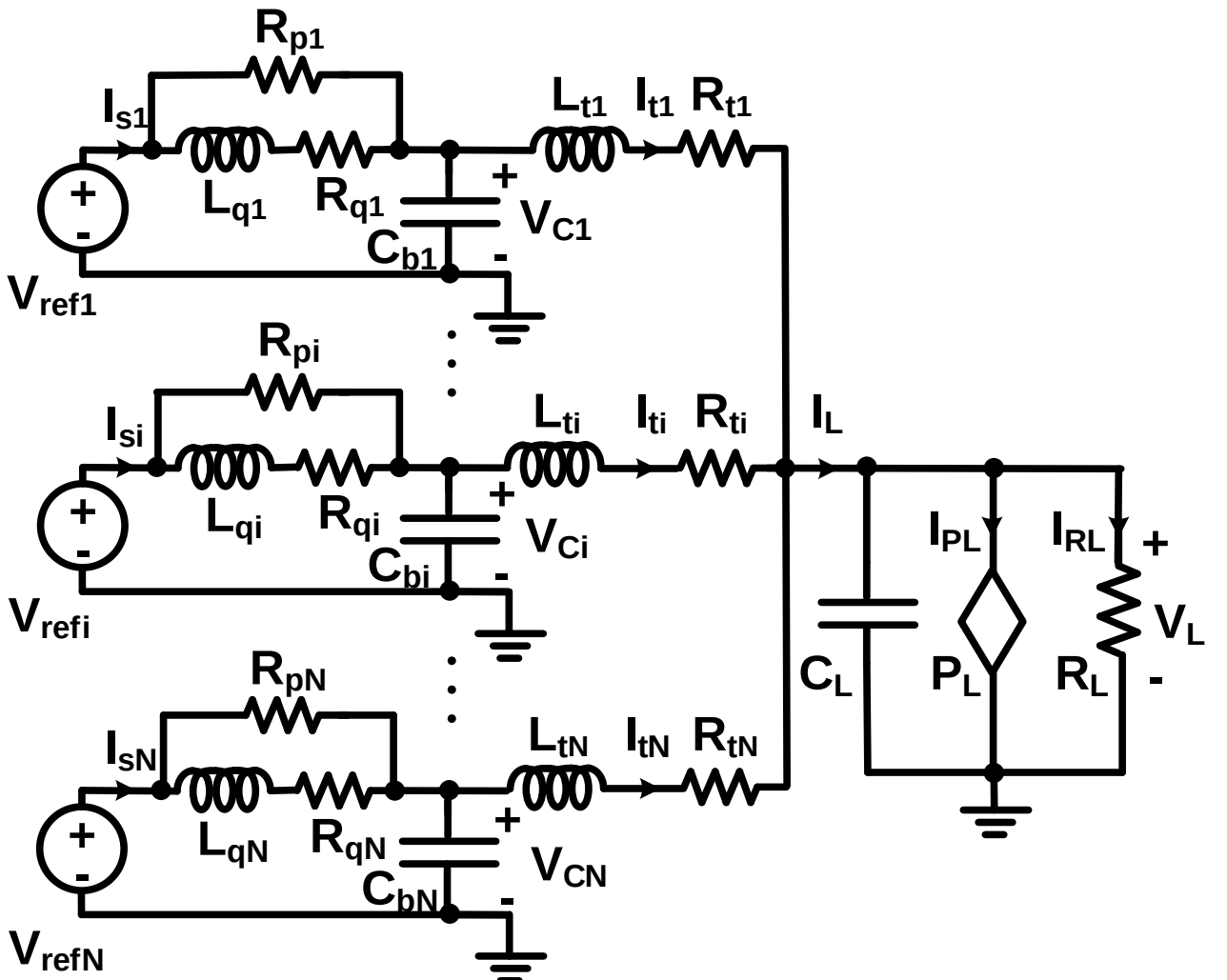

Fig. 4. The equivalent model of a DC microgrid with current-mode controllers.

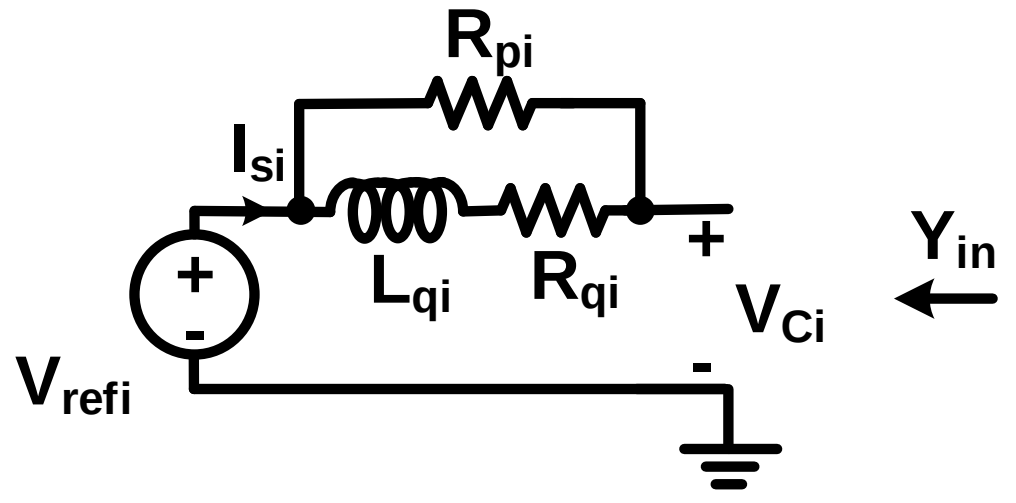

Fig. 5. The equivalent circuit of the proposed converter controller.

The DC microgrids with the novel proposed converter controllers have the following strengths compared to that with traditional droop controllers:

1) When the microgrid is in steady state, the equivalent impedance of the equivalent circuit of the proposed converter controllers (i.e. $1/Y_{in}$ in Fig. 5) is $\frac{R_{pi}R_{qi}}{R_{pi}+R_{qi}} \approx R_{qi}$ due to $R_{pi} \gg R_{qi}$. That is to say, the controller can be treated as a small resistor, leading to a small steady-state control error compared to that of the droop controller with large resistance, e.g. $R_{pi}$.

2) When the microgrid is in transient, the equivalent impedance of the equivalent circuit of the proposed controllers (i.e. $1/Y_{in}$ in Fig. 5) is $\frac{R_{pi}(sL_{qi}+R_{qi})}{R_{pi}+sL_{qi}+R_{qi}}$, which is nearly as large as $R_{pi}$ when $L_{qi}$ is set properly. It leads to the quick attenuation of energy at high frequencies, which is beneficial in maintaining the stability of the system. This characteristic makes the proposed controller superior over the traditional droop controller with small resistance, e.g. $R_{qi}$.

Besides the excellent performance of the novel converter controller, the similarity in the structure between the novel controller and a droop controller also makes it more convenient and promising to be developed in DC microgrids in practice. A simulation is carried out in section V. B to show the superiority of the novel proposed controller in detail.

## IV. Large-signal Stability Criteria in DC Microgrids with Closed-Loop Converter Controllers

### A. Introduction to the Potential of a Complete Circuit

**Definition 1 (Complete circuit [16])** A set of variables $i_1, \ldots, i_r, v_{r+1}, \ldots, v_{r+s}$ is called complete if they can be independent without leading to a violation of Kirchhoff's laws and if they determine at least one of the two variables, the current or the voltage, in each branch. A circuit is called complete if the set of variables $i_1, \ldots, i_r, v_{r+1}, \ldots, v_{r+s}$ is complete, where $i_1, \ldots, i_r$ denote the currents through inductors and $v_{r+1}, \ldots, v_{r+s}$ denote the voltage across capacitors.

**Definition 2 (Potential [21])** Define the potential function as follows:

$$P(i,v) = \sum_{\mu=r+1}^{s} v_\mu i_\mu \Big|_\Gamma + \sum_{\mu>r+s}^{b} \int_\Gamma v_\mu di_\mu \quad (3)$$

where $v_\mu$ and $i_\mu$ are the voltage and the current of the $\mu$-th element, respectively. Regarding the notations of the elements, $1,2,\ldots,r$ represent inductors; $r+1,\ldots,r+s$ represent capacitors; $r+s+1,\ldots,b$ represent nonlinear resistors and power sources. The integral term is also called current potential. The potential of some common elements in circuits is listed in the following table.

TABLE II
The Potential of Several Common Elements

| Capacitor | Linear resistor | Voltage source |
|---|---|---|
| $-v_\mu i_\mu$ | $-1/2{i_\mu}^2 R_\mu$ | $\int_\Gamma v_\mu di_\mu$ |

In the table, $R_\mu$ is the resistance of a linear resistor. The negative signs come from the generator convention. Besides, similar to current potential, the voltage potential is defined as follows:

$$q(i,v) = \sum_{\mu>r+s}^{b} \int_\Gamma i_\mu dv_\mu \quad (4)$$

A fundamental property of the potential of a circuit is that it only depends on the start point and endpoint of the chosen integral path whereas it is *independent of the path itself*, which is the same as gravitational potential energy. Moreover, there are two facts about the application of circuit potential:

1) It may not be practical to deal with the stability of a circuit with electronic elements with high nonlinearity using potential theory. Take the operational amplifier as an example—it consists of several highly nonlinear transistors, which causes difficulty in solving the equilibria of the system; besides, it has a large number of elements aside from the transistors, leading to a very complicated potential model. The potential analysis of the operational amplifier is questionable and misleading in [18].

2) Usually, the potential theory is applied to complete circuits. This does not imply that the potential function is not meaningful in an incomplete circuit, but sometimes it may not be interpreted as conveniently as that in a complete circuit. Normally, it is suggested to add capacitors in parallel and inductors in series to modify an incomplete circuit to a complete circuit. Then the original incomplete circuit can be treated as a limiting case of the modified complete circuit. The modification and limitation can be justified physically due to existing parasitic reactance in circuits.

### B. Sufficient Criteria for Global Asymptotic Stability in DC Microgrids with Closed-Loop Converter Controllers

As mentioned previously, the DC microgrids with the proposed closed-loop converter controllers are modeled as Fig. 4. Since it is an incomplete circuit, we add virtual inductors in series to modify it to a complete circuit. Suppose there is a virtual inductor $L_{pi}$, whose inductance is zero, in series with $R_{pi}$ in every converter controller. The modified model is shown in Fig. 6.

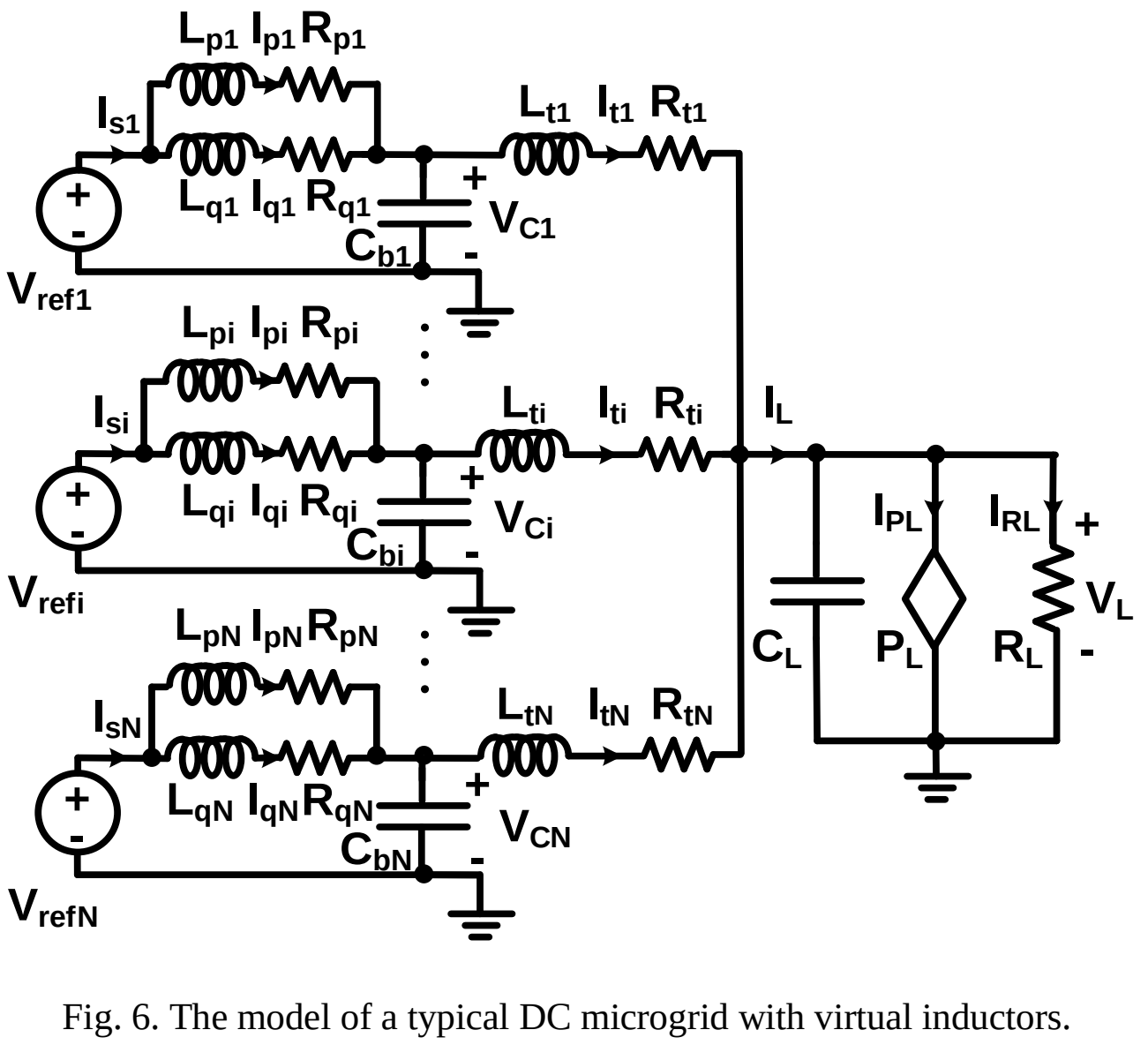

Fig. 6. The model of a typical DC microgrid with virtual inductors.

Here we consider a simplified CPL model as $\begin{cases} I_{PL} = I_{max}, V_L < V_{min} \\ V_L = P_L/I_{PL}, V_L \geq V_{min} \end{cases}$, where $I_{PL}$ and $V_L$ are the current and the output voltage of the CPL, separately. $P_L$ is the power of the CPL when $V_L \geq V_{min}$ . $V_{min}$ is the lower bound of the output voltage. $I_{max}$ is the upper bound of current.

The potential function of the system in Fig. 6 is written as

$$P(i,v) = \begin{cases} Z(i,v) + I_{max}(V_L - V_{min}) - P_L, & V_L < V_{min} \\ Z(i,v) + \int_{V_{min}}^{V_L} \frac{P_L}{v} dv - P_L, & V_L \geq V_{min} \end{cases} \quad (5)$$

where

$$Z(i,v) = \sum_{i=1}^{N} V_{refi}\left(I_{pi} + I_{qi}\right) - \frac{1}{2}\sum_{i=1}^{N} R_{pi}{I_{pi}}^2 - \frac{1}{2}\sum_{i=1}^{N} R_{qi}{I_{qi}}^2 - \frac{1}{2}\sum_{i=1}^{N} R_{ti}{I_{ti}}^2 - \sum_{i=1}^{N} V_{Ci}\left(I_{pi} + I_{qi} - I_{ti}\right) - \frac{{V_L}^2}{2R_L} - V_L\left(\sum_{i=1}^{N} I_{ti} - \frac{P_L}{V_L} - \frac{V_L}{R_L}\right) \quad (6)$$

The notations are corresponding to those marked in Fig. 6. The dynamic equation of the model in Fig. 6 is described as follows:

$$-J\frac{dx}{dt} = \frac{\partial P(x)}{\partial x} \quad (7)$$

where $x = [i \;\; v]^T$, $J = \begin{bmatrix} -L & 0 \\ 0 & C \end{bmatrix}$.

$i = [I_{p1}, I_{p2}, \dots, I_{pN}, I_{q1}, I_{q2}, \dots, I_{qN}, I_{t1}, I_{t2}, \dots, I_{tN}]$, $v = [V_{C1}, V_{C2}, \dots, V_{CN}, V_L]$, $L$ and $C$ are diagonal inductance matrix and diagonal capacitance matrix, respectively. Under this description, whether $J$ is positive definite is highly dependent on the values of $L$ and $C$. Therefore, we prefer to seek another expression of this system, which uses $(P^*, J^*)$ instead of $(P, J)$, such that

$$-J^*\frac{dx}{dt} = \frac{\partial P^*(x)}{\partial x} \quad (8)$$

where $J^*$ is always positive definite when the system is stable. Through equation transformation and superposition, the pair $(P^*, J^*)$ are obtained as follows:

$$J^* = \left(\lambda \mathbb{I} + \frac{\partial^2 P(x)}{\partial x^2} M\right) \cdot J, P^* = \lambda P + \frac{1}{2}\left(\frac{\partial P(x)}{\partial x}, M\frac{\partial P(x)}{\partial x}\right) \quad (9)$$

where $\mathbb{I}$ is an identity matrix, $M$ is a constant symmetric matrix, and $\lambda$ is a constant. The derivation of the pair $(P^*, J^*)$ is shown in Appendix A.

The following theorem is proposed to point out the sufficient conditions for global asymptotic stability in nonlinear circuit systems. The proof of Theorem 2 is shown in Appendix B.

**Theorem 2** Given a nonlinear circuit $\frac{dx}{dt} = f(x)$,

a) Let $P^*: \mathcal{R}^n \to \mathcal{R}$ be of the class $C^1$ such that:

i. $-J^*\frac{dx}{dt} = \frac{\partial P^*(x)}{\partial x}$ where $J^* \succ 0$

ii. $P^*(x)$ is radially unbounded, i.e., $P^*(x) \to \infty$ as $\|x\| \to \infty$

iii. $E := \{x \in \mathcal{R}^n | f(x) = 0\}$ , all equilibria of the nonlinear circuit are a compact set.

then every solution starting in $\mathcal{R}^n$ approaches $E$ as $t \to \infty$.

b) For those points on the set $E$ where $P^*$ is of class $C^2$, let $M = \{ x \in E | \frac{\partial^2 P^*}{\partial x^2} \geq 0\}$ , then every solution starting in $\mathcal{R}^n$ approaches $M$ as $t \to \infty$.

Theorem 2-a ensures that any trajectory of the system starting in $\mathcal{R}^n$ converges to the set $E$; however, it does not determine the stability of each equilibrium and cannot clarify which equilibrium the trajectory will converge to. Theorem 2-b not only determines the stability of every equilibrium but also shrinks the invariant set further. Next, we present the derivation of the large-signal sufficient criteria using Theorem 2. In this paper, we focus on the case where all equilibria of a microgrid satisfy $V_L \geq V_{min}$ hence the CPL operates as $V_L = P_L/I_{PL}$.

**Condition 0**: First we show $P^*: \mathcal{R}^n \to \mathcal{R}$ is of the class $C^1$.

a) $P(i,v)$ is continuous at $V = V_{min}$ because

$$\lim_{V \to V_{min}} \int_{V_{min}}^{V} \frac{P_L}{v} dv = \lim_{V \to V_{min}} I_{max}(V_L - V_{min}) = 0 \quad (10)$$

b) $\nabla P(i,v)$ is continuous at $V = V_{min}$ because

$$\left.\frac{\partial P}{\partial v}\right|_{v=v_{min}} = \frac{P_L}{V_{min}} = I_{max} \quad (11)$$

So $P(i,v)$ is of the class $C^1$. Choose $M = \begin{bmatrix} 2A^{-1} & 0 \\ 0 & 0 \end{bmatrix}$. Then it can be concluded that $P^*$ is also of the class $C^1$.

Second, it is verified that $P^*$ is of class $C^2$ on the set $E$ except for the operation point $(V_L, I_{PL}) = (V_{min}, I_{max})$, considering the characteristics of all circuit elements in our model.

**Condition 1**: $-J^*\frac{dx}{dt} = \frac{\partial P^*(x)}{\partial x}$ where $J^* \succ 0$. This condition is to ensure that the gradient of the potential function $P^*(x)$ is negative, i.e., $\dot{P^*}(x) = \frac{\partial P^*(x)}{\partial x} \cdot \frac{dx}{dt} < 0$. It guarantees that state variable $x$ goes along the direction in which $P^*(x)$ decreases. Based on this condition, we derive the first condition for global asymptotic stability shown as follows.

$$\sigma_{max}\left(L^{1/2} A^{-1} \gamma C^{-1/2}\right) < 1 \quad (12)$$

The derivation can be found in Appendix C.

**Condition 2**: $P^*(x)$ is radially unbounded, i.e., $P^*(x) \to \infty$ as $\|x\| \to \infty$.

This condition will be checked directly in specific circuits.

**Condition 3**: $E := \{x \in \mathcal{R}^n | f(x) = 0\}$, all equilibria of the nonlinear circuit form a compact set.

Consider the system in equation (8) again. We note that the equilibria of the system are exactly the stationary points of $P^*(x)$, i.e., $\partial P^*(x)/\partial x = 0$. Since the number of the equilibria in a circuit system is finite, $E$ is a compact set.

**Condition 4**: Solve $M = \{ x \in E | \frac{\partial^2 P^*}{\partial x^2} \geq 0\}$. This condition is deployed to distinguish between stable equilibria and unstable ones. Then we solve this condition in detail:

$$\left.\frac{\partial^2 P^*(x)}{\partial x^2}\right|_{x=x_e} \geq 0 \quad (13)$$

where $x_e = (i_e, v_e)$ are equilibria in a microgrid.

Rewrite $P(x)$ ( $V_L \geq V_{min}$) in equation (5) in this form:

$$P(i,v) = -\frac{1}{2}(i, Ai) + B(v) + (i, \gamma v - \alpha) \tag{14}$$

where $i = [I_{p1}, I_{p2}, \dots, I_{pN}, I_{q1}, I_{q2}, \dots, I_{qN}, I_{t1}, I_{t2}, \dots, I_{tN}]_{3N\times1}$, $v = [V_{C1}, V_{C2}, \dots, V_{CN}, V_L]_{(N+1)\times1}$,

$$\gamma = \begin{bmatrix} -\mathbb{I}_{N\times N} & 0_{N\times1} \\ -\mathbb{I}_{N\times N} & 0_{N\times1} \\ \mathbb{I}_{N\times N} & -1_{N\times1} \end{bmatrix}_{(3N)\times(N+1)},$$

$A = diag[R_{p1}, \dots, R_{pN}, R_{q1}, \dots, R_{qN}, R_{t1}, \dots R_{tN}] = diag[R_p, R_q, R_t]$.

According to equation (9) we have

$$P^* = \lambda P + \frac{1}{2}\left(\frac{\partial P(x)}{\partial x}, M\frac{\partial P(x)}{\partial x}\right) \tag{15}$$

Therefore, given $\lambda = 1$, $M = \begin{bmatrix} 2A^{-1} & 0 \\ 0 & 0 \end{bmatrix}$, we have

$$\left.\frac{\partial^2 P^*(x)}{\partial x^2}\right|_{x=x_e} = \left.\frac{\partial^2 P(x)}{\partial x^2} + \frac{d\left(\frac{\partial^2 P(x)}{\partial x^2} M \frac{\partial P(x)}{\partial x}\right)}{dx}\right|_{x=x_e} \tag{16}$$

Since

$$\left.\frac{\partial^2 P(x)}{\partial x^2}\right|_{x=x_e} = \left.\begin{bmatrix} -A & \gamma \\ \gamma^T & \frac{\partial^2 B(v)}{\partial v^2} \end{bmatrix}\right|_{v=v_e} \tag{17}$$

$$\left.\frac{\partial P(x)}{\partial x}\right|_{x=x_e} = \left.\begin{bmatrix} -Ai + \gamma v - \alpha \\ \frac{\partial B(v)}{\partial v} + \gamma^T i \end{bmatrix}\right|_{v=v_e} \tag{18}$$

where $x = [i\ v]^T$ and $v_e$ notates $v$ in steady state, then equation (13) is calculated as:

$$\left.\frac{\partial^2 P^*(x)}{\partial x^2}\right|_{x=x_e} = \left.\begin{bmatrix} A & -\gamma \\ -\gamma^T & \frac{\partial^2 B(v)}{\partial v^2} + 2\gamma^T A^{-1}\gamma \end{bmatrix}\right|_{v=v_e} \geq 0 \tag{19}$$

According to the Schur complement condition for positive semi-definiteness, if $A \succ 0$, $X$ is positive semi-definite if and only if $X/A$ is positive semi-definite, where $X$ is a symmetric matrix given by $X = \begin{bmatrix} A & B \\ B^T & C \end{bmatrix}$, $X/A = C - B^T A^{-1} B$ is the Schur complement of $A$.

In equation (19), we know $A = diag([R_p, R_q, R_t]) \succ 0$. Hence, equation (19) can be converted as follows:

$$\left.\frac{\partial^2 B(v)}{\partial v^2} + \gamma^T A^{-1}\gamma\right|_{v=v_e} \geq 0 \tag{20}$$

That is to say, condition (13) can be solved by

$$\left.\frac{\partial^2 B(v)}{\partial v^2} + \gamma^T A^{-1}\gamma\right|_{v=v_e} \geq 0 \tag{21}$$

Considering that

$$\left.\frac{\partial^2 B(v)}{\partial v^2}\right|_{v=v_e} = \begin{bmatrix} 0_{N\times N} & 0_{N\times1} \\ 0_{1\times N} & \frac{1}{R_L} - \frac{P_L}{{v_e}^2} \end{bmatrix} \tag{22}$$

$$\gamma^T A^{-1}\gamma|_{v=v_e} = \begin{bmatrix} \ddots & & & \vdots \\ & \frac{1}{R_{pi}} + \frac{1}{R_{qi}} + \frac{1}{R_{ti}} & & -\frac{1}{R_{ti}} \\ & & \ddots & \vdots \\ \cdots & -\frac{1}{R_{ti}} & \cdots & \sum_{i=1}^{N}\frac{1}{R_{ti}} \end{bmatrix} \tag{23}$$

We notate $U = diag[\frac{1}{R_{p1}} + \frac{1}{R_{q1}} + \frac{1}{R_{t1}}, \dots, \frac{1}{R_{pN}} + \frac{1}{R_{qN}} + \frac{1}{R_{tN}}]$, $V = \begin{bmatrix} -\frac{1}{R_{t1}} & \cdots & -\frac{1}{R_{tN}} \end{bmatrix}$, $W = \frac{1}{R_L} - \frac{P_L}{{v_e}^2} + \sum_{i=1}^{N}\frac{1}{R_{ti}}$.

Then we have

$$\left.\frac{\partial^2 B(v)}{\partial v^2} + \gamma^T A^{-1}\gamma\right|_{v=v_e} = \begin{bmatrix} U & V^T \\ V & W \end{bmatrix} \tag{24}$$

Use the Schur complement for positive semi-definiteness: since $R_{p1}, \dots, R_{pN}, R_{q1}, \dots, R_{qN}, R_{t1}, \dots, R_{tN}$ are all positive, $U \succ 0$ holds. Therefore, we have

$$\left.\frac{\partial^2 B(v)}{\partial v^2} + \gamma^T A^{-1}\gamma\right|_{v=v_e} \geq 0 \Leftrightarrow \det(W - VU^{-1}V^T) \geq 0 \tag{25}$$

Considering that

$$\det(W - VU^{-1}V^T) = W - \sum_{i=1}^{N}\frac{1}{{R_{ti}}^2\left(\frac{1}{R_{pi}} + \frac{1}{R_{qi}} + \frac{1}{R_{ti}}\right)} \geq 0 \tag{26}$$

condition (13) is solved by

$$W - \sum_{i=1}^{N}\frac{1}{{R_{ti}}^2\left(\frac{1}{R_{pi}} + \frac{1}{R_{qi}} + \frac{1}{R_{ti}}\right)} \geq 0 \tag{27}$$

where $W = \frac{1}{R_L} - \frac{P_L}{{v_e}^2} + \sum_{i=1}^{N}\frac{1}{R_{ti}}$.

In conclusion, the sufficient criteria for global asymptotic stability of a DC microgrid with distributed closed-loop converter controllers are shown as follows:

a). $P^*(x)$ is radially unbounded, i.e.

$$P^*(x) \to \infty \text{ as } \|x\| \to \infty. \tag{28}$$

b).

$$\begin{cases} \sigma_{max}(L^{1/2}A^{-1}\gamma C^{-1/2}) < 1 \\ W - \sum_{i=1}^{N}\frac{1}{{R_{ti}}^2\left(\frac{1}{R_{pi}} + \frac{1}{R_{qi}} + \frac{1}{R_{ti}}\right)} \geq 0 \end{cases} \tag{29}$$

where $\sigma_{max}(\cdot)$ is the largest singular value, $W = \frac{1}{R_L} - \frac{P_L}{{v_e}^2} + \sum_{i=1}^{N}\frac{1}{R_{ti}}$.

To illustrate the difference between our proposed large-signal stability criteria and that proposed in Brayton-Moser's theory [16], the comparison results are listed in Table III. It is seen from the table that Brayton-Moser's theory does not consider condition 0, condition 3, and condition 4. Generally, it shows two defects: first, it ignores that $P^*$ should be defined as $P^*: \mathcal{R}^n \to \mathcal{R}$ and be of the class $C^1, C^2$. Considering the most common CPL model which only considers the hyperbolic operation function as $V = P/I (V > 0, I > 0)$, it is not defined at $V = 0$ or $I = 0$, which means Brayton-Moser's theory actually cannot work on it. Besides, $\frac{\partial^2 P^*(x)}{\partial x^2}$ will not exist if $P^*$ is not guaranteed to be of the class $C^1, C^2$. Second, Brayton-Moser's theory only determines the sufficient conditions for the convergence to the set $E$ which includes all equilibria; however, it does not indicate which equilibrium the system will converge to. In the real operation of a microgrid, it is critical to clarify the stability of every equilibrium point and to ensure that the system converges to the expected equilibrium. Condition 4 proposed in our method solves this issue. The system will converge only to where condition 4 is satisfied.

TABLE III
COMPARISON BETWEEN DIFFERENT STABILITY CRITERIA

| Conditions considered | The novel proposed method | Brayton-Moser's method |
|---|---|---|
| Condition 0: $P^*$ is of class $C^1$; $P^*$ is of class $C^2$ on the set $E$ except for the point $(V_L, I_{PL}) = (V_{min}, I_{max})$ | ✓ | × |
| Condition 1: $J^* \succ 0$ | ✓ | ✓ |
| Condition 2: $P^*$ is radially unbounded | ✓ | ✓ |
| Condition 3: all equilibrium points of the system form a compact set | ✓ | × |
| Condition 4: $\left.\frac{\partial^2 P^*(x)}{\partial x^2}\right\|_{x=x_e} \geq 0$ | ✓ | × |

Although the above large-signal stability criteria are proposed for the DC microgrids with closed-loop converter controllers, it also can be tailored to fit for the DC microgrids with open-loop converter controllers. Compared to the closed-loop converter controller, the open-loop converter controller refers to no feedback control between converter parameters and operation parameters of a DC microgrid. The derivation is skipped here.

## V. CASE STUDY

The case study consists of four parts, which correspondingly verify our four contributions indicated in the introduction of this paper. The above derivation obtains the sufficient conditions for large-signal stability in a DC microgrid, which benefits the design and operation of a stable DC microgrid. The main steps of the implementation of the proposed stability analysis in practice are shown as follows.

**Algorithm 1:** The Novel Proposed Stability Analysis of a DC microgrid in Practice

Step 1: Extract a circuit model from a practical DC microgrid
Step 2: Calculate the potential function of the circuit model
Step 3: Solve the proposed potential-based sufficient conditions for large-signal stability using equation (28)(29)
Step 4: Obtain the ranges of microgrid parameters for the global stability of the DC microgrid

### *A. Verification of the Proposed Large-Signal Stability Criteria in DC Microgrids*

A simulation model is built as depicted in Fig. 7 to verify the correctness of our proposed stability criteria. The simulation parameters are set as shown in Table IV. Here, we explore the performance of the proposed stability criteria through the sensitivity analysis of the power of CPL $P_L$. It is determined that there exists a theoretical stability boundary around $P_L = 805W$ using our proposed stability criteria. A checklist of parts of data points is shown in Table V. Then we test these data points using Matlab/Simulink to show the correspondence between the simulation results and the theoretical results derived from our proposed stability criteria. The voltage at PoL is measured to reflect the stability of the system, as shown in Fig. 8.

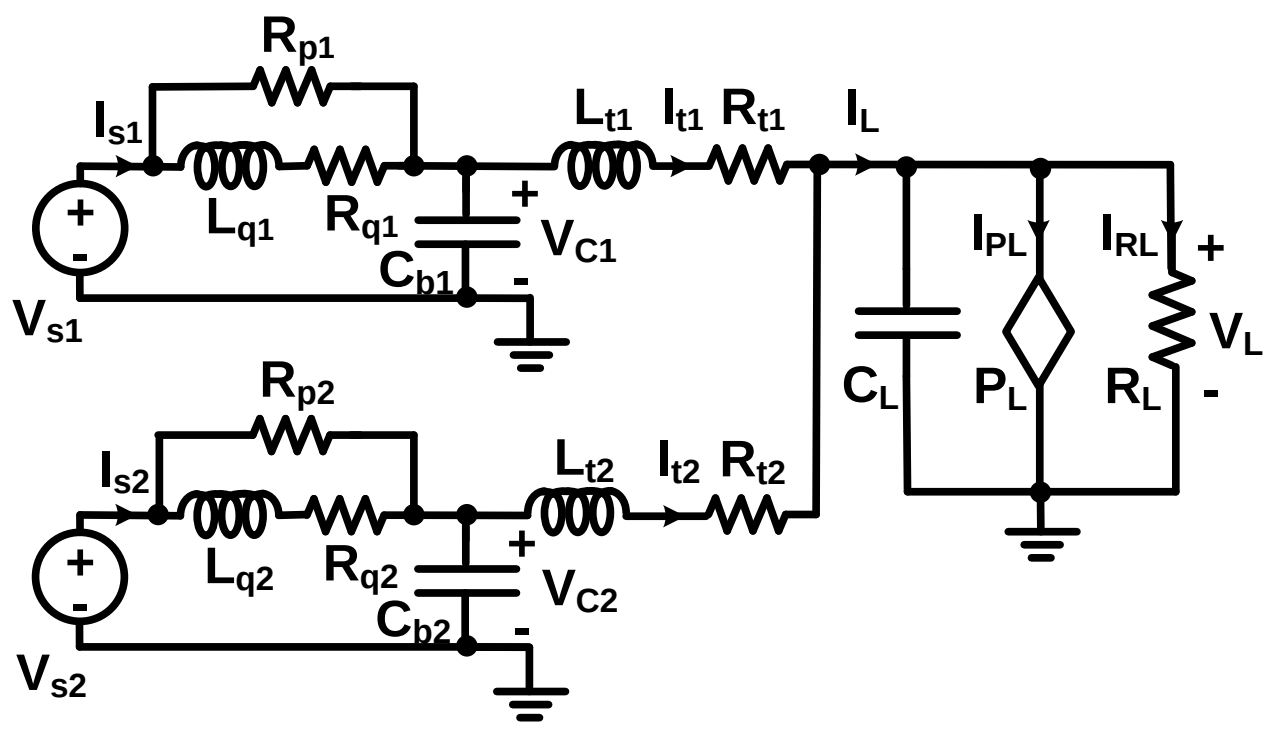

Fig. 7. The simulation model of a DC microgrid with distributed converter controllers.

TABLE IV
SIMULATION PARAMETERS
(The unit: V, H, F, Ohm, W)

| | | | | | | | | | |
|---|---|---|---|---|---|---|---|---|---|
| $L_{q1}$ | 1.0 | $L_{t1}$ | 0.5 | $L_{q2}$ | 1.0 | $L_{t2}$ | 0.5 | $C_L$ | 1.0 |
| $R_{q1}$ | 0.9 | $R_{t1}$ | 3.0 | $R_{q2}$ | 0.9 | $R_{t2}$ | 3.0 | $P_L$ | 800 |
| $R_{p1}$ | 0.6 | $C_{b1}$ | 5.0 | $R_{p2}$ | 0.6 | $C_{b2}$ | 5.0 | $R_L$ | 2.0 |
| $V_{s1}$ | 100 | $V_{s2}$ | 100 | | | | | | |

TABLE V
THE CHECKLIST OF TEST DATA

| $P_L$(W) | 800 | 805 | 810 | 825 |
|---|---|---|---|---|
| Check: whether the stability criteria (28)(29) are satisfied by the system parameters (Yes/No) | Yes | Yes | No | No |

The system starts to operate at $t = 0s$. It is observed from Fig. 8 that the voltage at PoL rises quickly from 0V to the steady-state value (about 55V) and then keeps stable until the CPL is plugged into the system at $t = 20s$. After the CPL is plugged in, the system maintains stability when $P_L = 800W$ and $P_L = 805W$, whereas it oscillates severely when $P_L = 825W$. Notably, the system is instable but very approaching the stable state when $P_L = 810W$, which is nearby the critical state. The simulation results completely correspond to the theoretical results in Table V, which verifies the correctness of our proposed stability criteria.

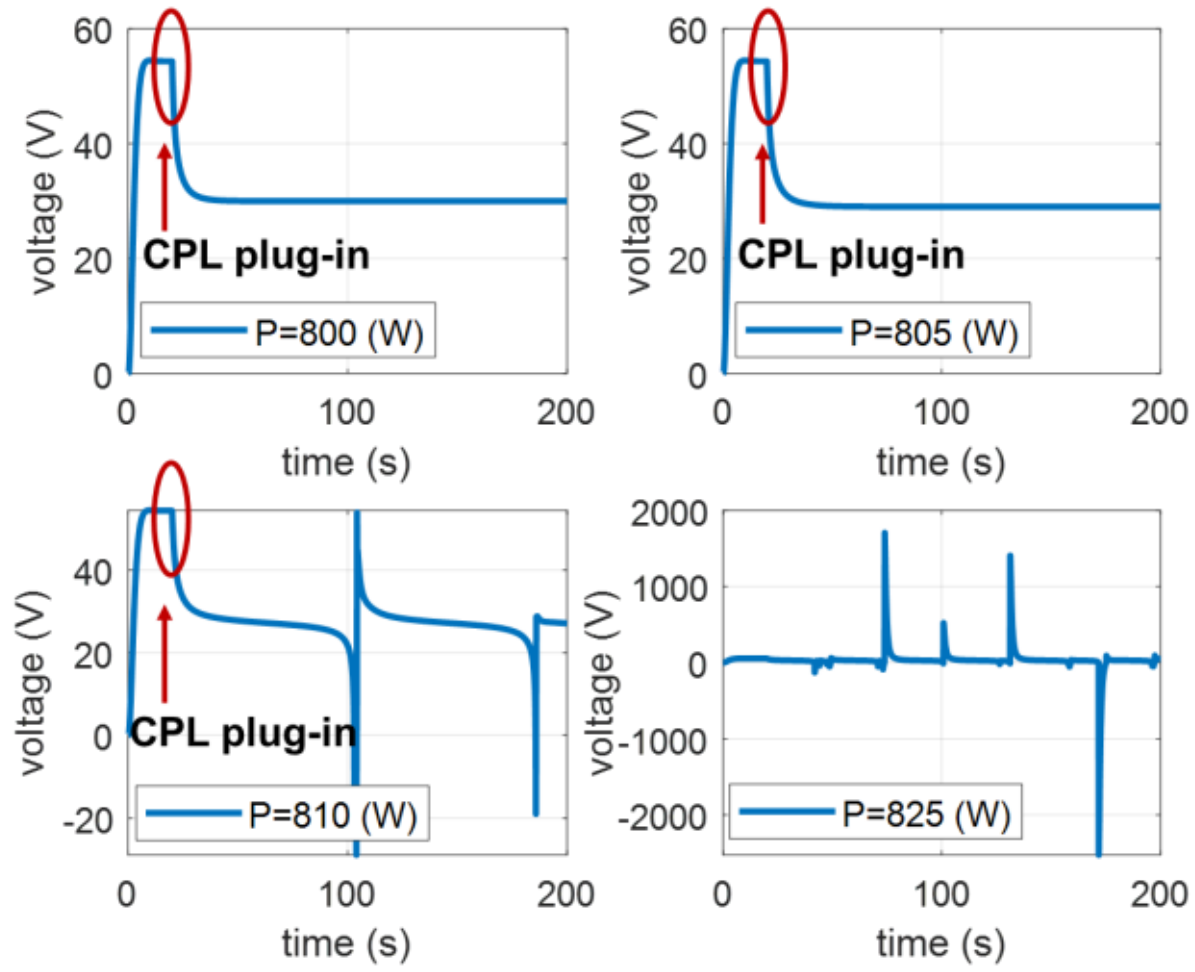

Fig. 8. The voltage at PoL in the model with different power of CPL.

The above simulation is based on the *averaging model* of power converters. Next, we show another example with the *switching model* of power converters to make the discussion more convincing and comprehensive. The diagram of the simulation is shown in Fig. 9 and the simulation parameters are as shown in Table VI. The simulation platform is PLECS. The system starts to operate at $t = 0s$, and the CPL is plugged in at $t = 3s$.

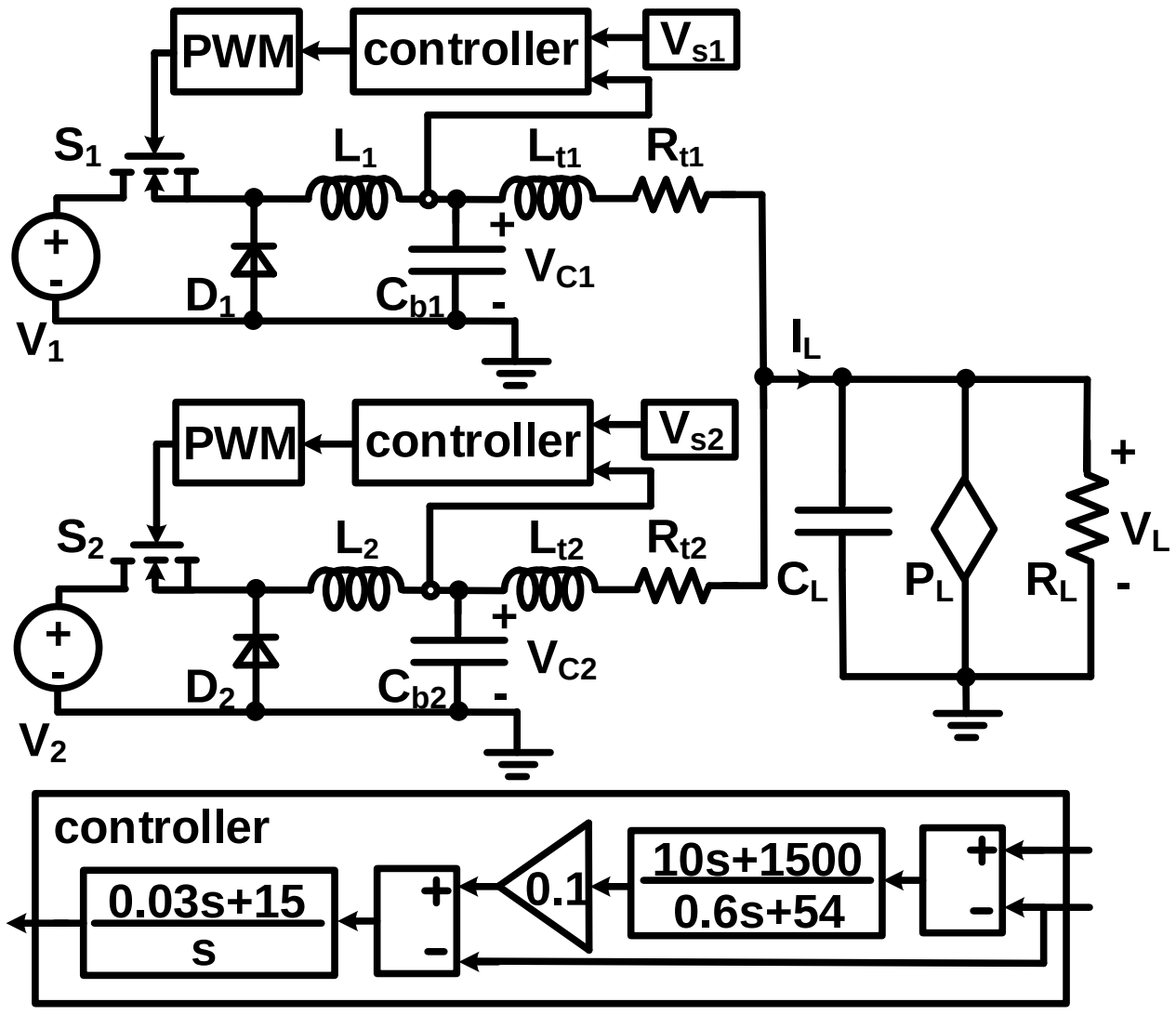

Fig. 9. The simulation diagram of a DC microgrid with the switching model of converters.

TABLE VI
SIMULATION PARAMETERS
(The unit: V, H, F, Ohm, W)

| | | | | | | | | | |
|---|---|---|---|---|---|---|---|---|---|
| $V_1$ | 200 | $L_1$ | $10^{-3}$ | $C_{b1}$ | 0.1 | $L_{t1}$ | 0.5 | $R_{t1}$ | 4.0 |
| $V_2$ | 200 | $L_2$ | $10^{-3}$ | $C_{b2}$ | 0.1 | $L_{t2}$ | 0.5 | $R_{t2}$ | 4.0 |
| $V_{s1}$ | 100 | $V_{s2}$ | 100 | $C_L$ | 0.1 | $P_L$ | 500 | $R_L$ | 20 |

On one hand, it is theoretically verified that the system is stable since the proposed stability criteria are satisfied by the simulation parameters in Table VI. On the other hand, it can be concluded from the simulation results that the system is stable after going through startup and the plug-in of CPL. The dynamic responses of the load current $I_L$ and the voltage at PoL $V_L$ are shown in Fig. 10 (a).

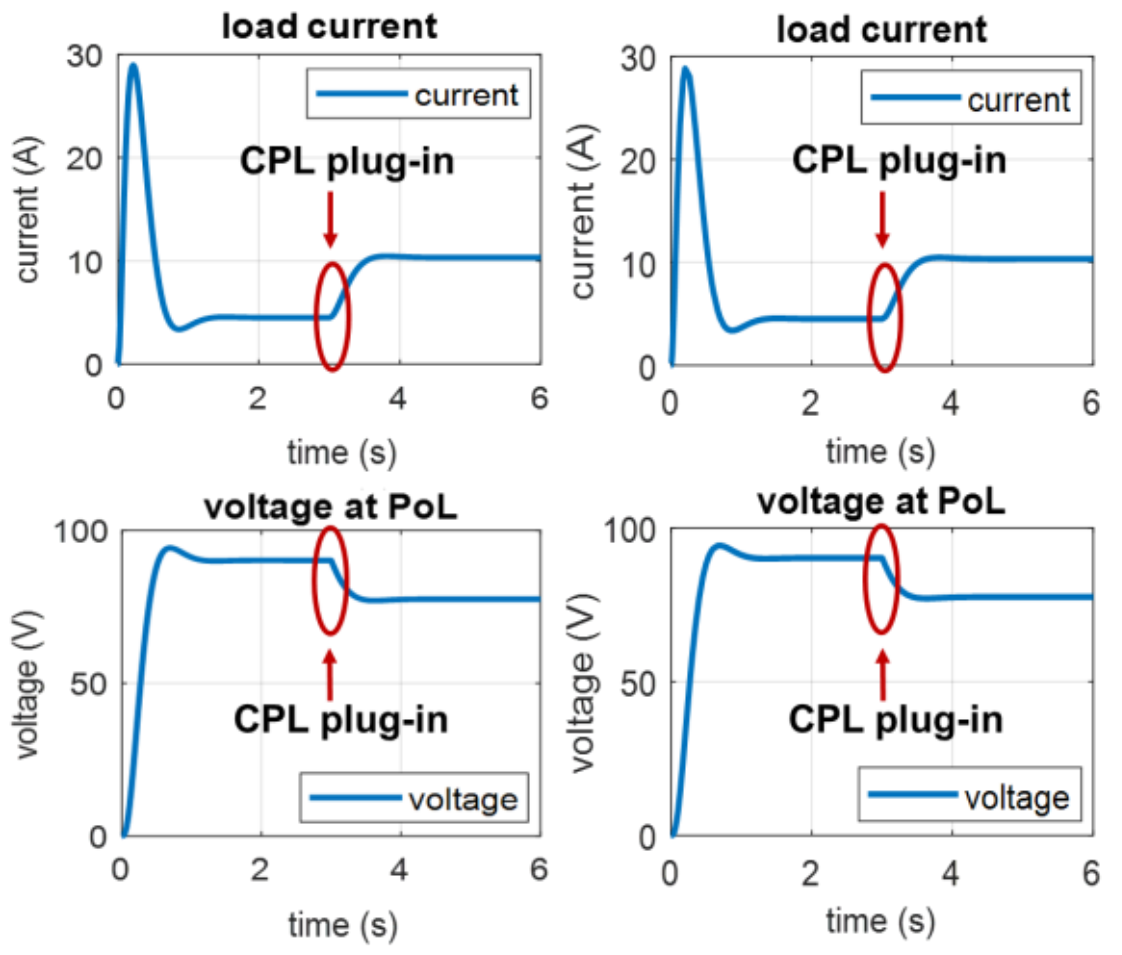

(a) (b)

Fig. 10. (a) The dynamic responses of the system with the switching model of converters; (b) The dynamic responses of the system with the averaging model of converters.

Moreover, we compare the dynamic responses of the system using the switching model in Fig. 9 and that using the averaging model. The diagram and the simulation parameters of the system using the averaging model are shown in Fig. 7 and Table VII, respectively. Figure 10 (b) presents the dynamic responses of the system using the averaging model, which exhibits an excellent agreement with that using the switching model under much smaller computational complexities. It can be concluded that it is reasonable to employ the averaging model instead of the switching model to simplify the model of a DC microgrid.

TABLE VII
SIMULATION PARAMETERS
(The unit: V, H, F, Ohm, W)

| | | | | | | | | | |
|---|---|---|---|---|---|---|---|---|---|
| $L_{q1}$ | 0.01 | $L_{t1}$ | 0.5 | $L_{q2}$ | 0.01 | $L_{t2}$ | 0.5 | $C_L$ | 0.1 |
| $R_{q1}$ | 0.9 | $R_{t1}$ | 4.0 | $R_{q2}$ | 0.9 | $R_{t2}$ | 4.0 | $R_L$ | 20 |
| $R_{p1}$ | 0.6 | $C_{b1}$ | 0.1 | $R_{p2}$ | 0.6 | $C_{b2}$ | 0.1 | $P_L$ | 500 |
| $V_{s1}$ | 100 | $V_{s2}$ | 100 | | | | | | |

*B. The Superiority of the Novel Proposed Converter Controllers*

In this section, a MATLAB/Simulink-based model of Fig. 4 is built to demonstrate the superiority of the novel proposed closed-loop converter controllers. Here, we choose a traditional droop controller as a benchmark. The simulation results of the stability analysis of a small-scale microgrid with different controllers are presented.

The simulation model is built as depicted in Fig. 11. In two different simulation scenarios, we deploy the novel proposed controllers and droop controllers in our model separately. The steady-state circuit of the model with the proposed controllers is kept equivalent to that with droop controllers. The CPL is plugged in at $t = 5s$. The simulation parameters are shown in the following table. The dynamic responses of the model using different controllers are shown in Fig. 12.

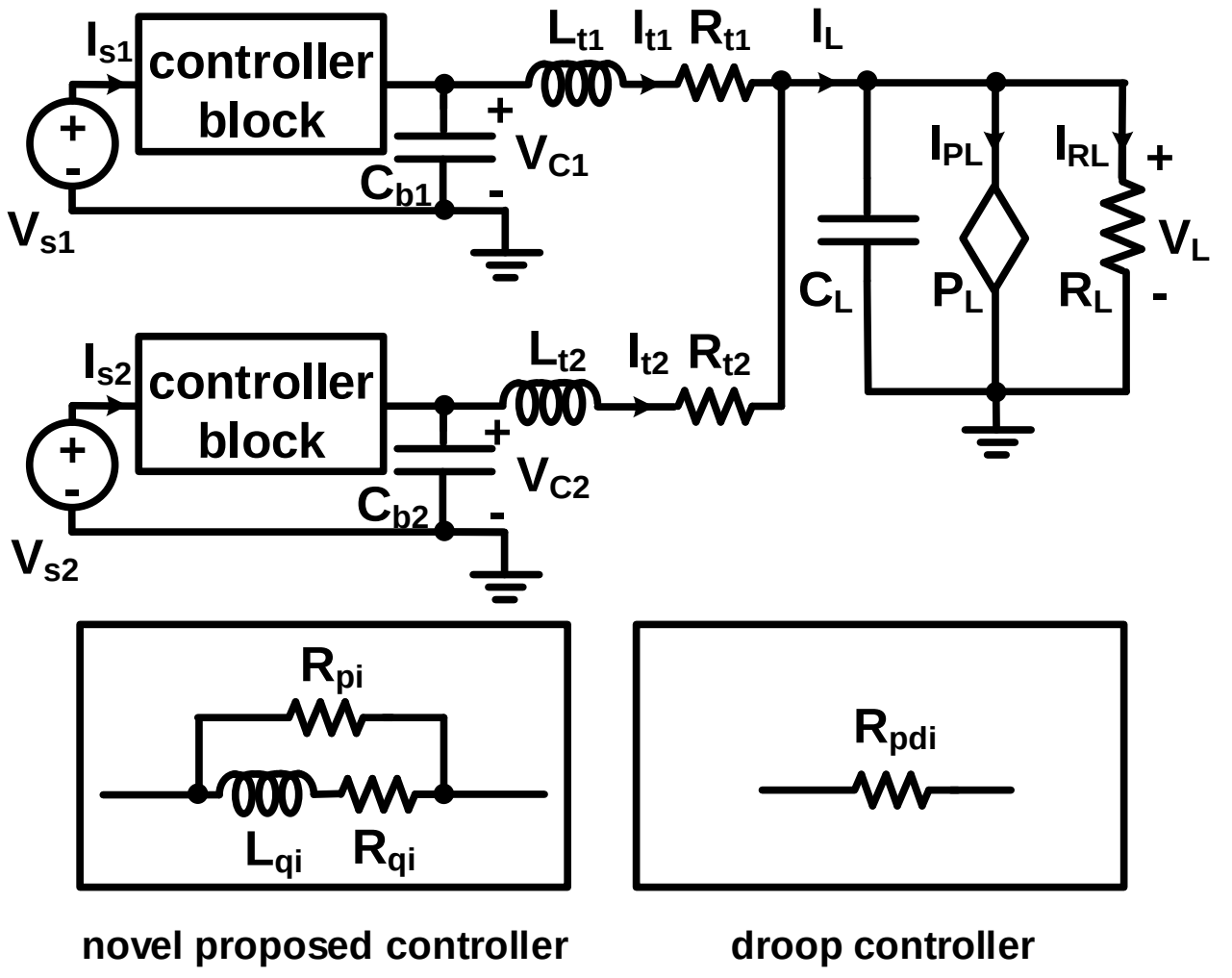

Fig. 11. The simulation model of a DC microgrid with different controllers.

TABLE VIII
SIMULATION PARAMETERS
(The unit: V, H, F, Ohm, W)

| | | | | | | | | | |
|---|---|---|---|---|---|---|---|---|---|
| $L_{q1}$ | 2.0 | $L_{t1}$ | 0.5 | $L_{q2}$ | 2.0 | $L_{t2}$ | 0.5 | $C_L$ | 0.05 |
| $R_{q1}$ | 1.25 | $R_{t1}$ | 0.01 | $R_{q2}$ | 1.25 | $R_{t2}$ | 0.01 | $R_L$ | 10 |
| $R_{p1}$ | 5.0 | $C_{b1}$ | 0.01 | $R_{p2}$ | 5.0 | $C_{b2}$ | 0.01 | $P_L$ | 530 |
| $V_{s1}$ | 100 | $V_{s2}$ | 100 | $R_{pd1}$ | 1.0 | $R_{pd2}$ | 1.0 | | |

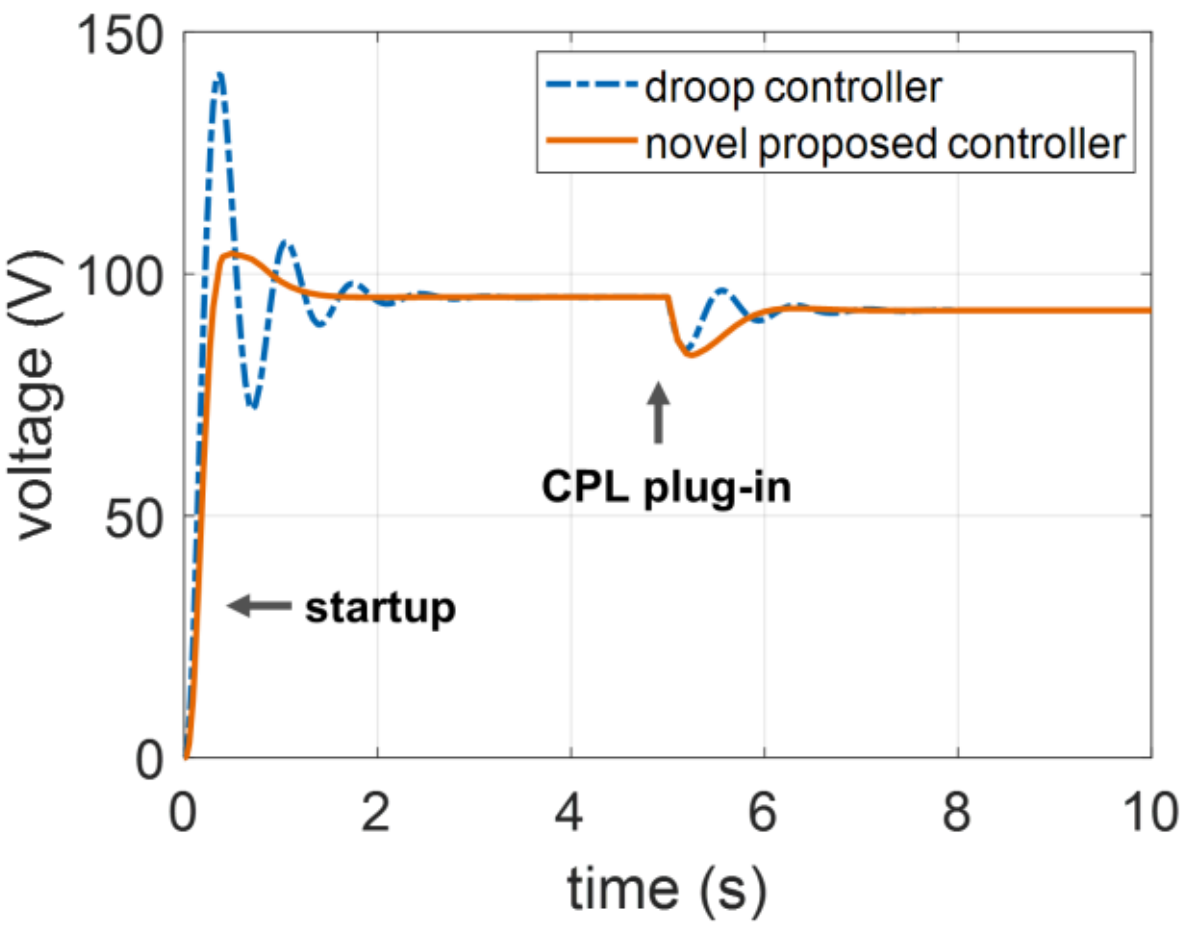

Fig. 12. The dynamic response of the voltage at PoL in the microgrid model.

First, during the startup of the system, it is observed that the novel proposed controller has quite a lower overshoot than the droop controller, which improves the stability of the system during its startup. Comparatively, the droop controller is not stable until going through more than three cycles, leading to severe oscillation. Second, when the CPL is plugged in ($t = 5\ s$), the voltage drops suddenly to around 78V. During the next seconds, the voltage is going back to about 92V with the help of different controllers. Compared to the oscillation caused by the droop controller, the novel proposed controller realizes a smoother dynamic response, a smaller overshoot, and smaller deviations in the procedure from 78V to 92V, which shows the superiority of the novel proposed controller.

In conclusion, it can be seen from the simulation that the novel proposed controller ensures a smaller voltage overshoot and smaller voltage deviations than the droop controller, which acquires a smoother and more stable dynamic response of the voltage at PoL. The proposed controller successfully overcomes the dilemma of a traditional droop controller, which has to balance the tradeoff between a large overshoot and large steady-state errors.

## *C. Defects of Brayton-Moser's Mixed Potential Theory*

In this section, we present an example where Brayton-Moser's mixed potential theory [16] cannot obtain sufficient criteria for the stability of nonlinear circuits, using a second-order RLC circuit as depicted in Fig. 13. Suppose $R_L$ is a constant negative resistor, i.e., $R_L < 0$. On one hand, the *constant* negative resistor $R_L$ is different from the property of the CPL model, since the CPL model is equivalent to an *incremental* negative resistor; on the other hand, the *constant* negative resistor $R_L$ has similar characteristics to the CPL—probably leading to the instability of a circuit. The advantage of this model is that it presents similar characteristics to nonlinear circuits in terms of instability with lower computational costs.

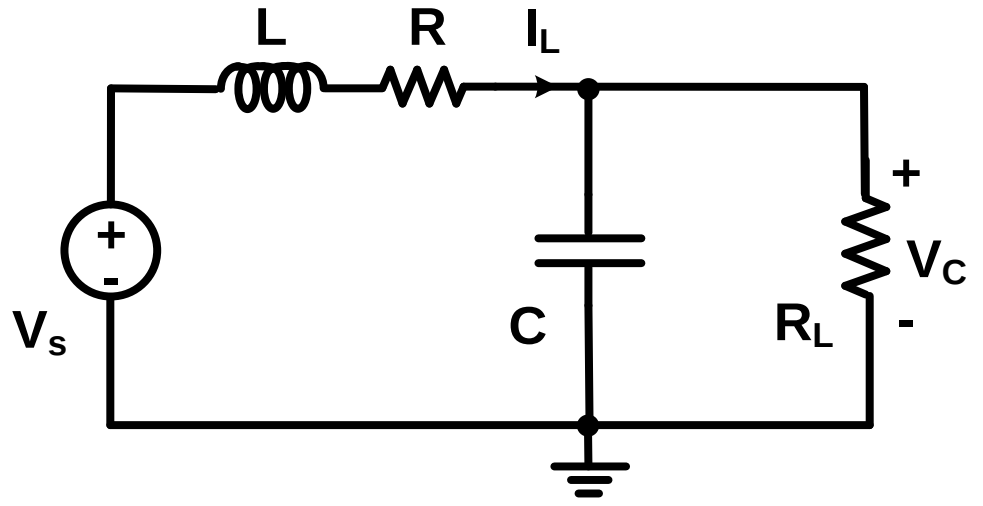

Fig. 13. The model of a second-order RLC circuit.

First of all, we solve the stability region of the circuit in Fig. 13 using a classic method based on the root analysis of the transfer function. The purpose of this step is to provide a correct stability region as a benchmark. Although this classic method is often utilized to obtain the small-signal stability region, it is applicable to determine the large-signal stability region for *linear systems*. In fact, the small-signal stability region is the same as the large-signal stability region in *linear systems*. The circuit model in Fig. 13 is a linear system with no doubt.

The transfer function $H(s)$ of the circuit model is as follows:

$$H(s) = \frac{I_L}{V_s} = \frac{1}{R + sL + \frac{1/sC \cdot R_L}{1/sC + R_L}} = \frac{CR_L(s + 1/CR_L)}{s^2LCR_L + s(L + CR_LR) + R_L + R} \tag{30}$$

The sufficient criteria for the stability of circuits are that both poles of the transfer function have non-positive real parts (the two poles cannot be zero at the same time). Since $R_L < 0$, according to the characteristics of the quadratic function, we obtain

$$real(s_1) < 0, real(s_2) < 0 \ \Rightarrow \begin{cases} L + CR_LR < 0 \\ R_L + R < 0 \end{cases} \tag{31}$$

Then we obtain the stability region as follows:

$$\frac{L}{C|R_L|} < R < |R_L| \tag{32}$$

This result is treated as a benchmark to show the defects in Brayton-Moser's mixed potential theory.

Next, we make a comparison between the stability criteria derived from Brayton-Moser's potential theory and that derived from our proposed criteria, which is shown in Table IX. The derivation of the stability criteria is in Appendix D.

TABLE IX
STABILITY CRITERIA USING DIFFERENT METHODS

| Method | Stability region | |
|---|---|---|
| Root analysis (benchmark) | $\frac{L}{C\|R_L\|} < R < \|R_L\|$ | (33) |
| Brayton-Moser's theory | $\{\phi\}$ | (34) |
| The novel proposed criteria | $\frac{L}{C\|R_L\|} < R < \|R_L\|$ | (35) |

where $\{\phi\}$ represents the empty set. At the beginning of section V.C, we explain the reason that the root analysis is utilized as a benchmark in large-signal stability analysis. It can be seen from equation (34) that Brayton-Moser's mixed potential theory cannot provide sufficient conditions for large-signal stability, even for a typical linear second-order RLC circuit. Therefore, we conclude that Brayton-Moser's theory cannot solve stability criteria in nonlinear circuits, such as the circuits with CPLs. Comparatively, condition (35) is the same as the result solved in the complex frequency domain.

In conclusion, Brayton-Moser's mixed potential theory cannot obtain sufficient criteria for stability in linear circuits and nonlinear circuits. Besides, considering this illustrative example and the simulation in section V. A, it is demonstrated that the novel proposed method solves the stability criteria rigorously and works well on both linear circuits and nonlinear circuits.

### *D. The Superiority of a Large-Signal Stability Region over a Small-Signal Stability Region*

In this section, we compare the large-signal stability region and the small-signal stability region of a microgrid model to demonstrate the significance of large-signal stability and its superiority over small-signal stability, taking the model shown in Fig. 4 as an example.

First, we formulate the small-signal stability analysis of the model in Fig. 4. Notate the current of the inductor $L_{qi}$ by $I_{qi}$; other notations are as marked in Fig. 4. Then the dynamic equations $F(\dots I_{qi}\dots,\dots I_{ti}\dots,\dots V_{Ci}\dots,V_L)$ of the microgrid model in Fig. 4 are as follows:

$$\begin{cases} L_{qi}\dfrac{dI_{qi}}{dt} = -I_{qi}R_{qi} + V_{refi} - V_{Ci} \\ C_{bi}\dfrac{dV_{Ci}}{dt} = \dfrac{V_{refi} - V_{Ci}}{R_{Pi}} + I_{qi} - I_{ti} \\ V_{Ci} - V_L = L_{ti}\dfrac{dI_{ti}}{dt} + R_{ti}I_{ti} \\ C_L\dfrac{dV_L}{dt} = \sum_{i=1}^{N} I_{ti} - \dfrac{P_L}{V_L} - \dfrac{V_L}{R_L} \end{cases} \tag{36}$$

We notate the steady-state voltage at PoL by $V_L{}^*$, which is solved by the following equations:

$$\begin{cases} V_{refi} - I_{si}R_{eqi} = \dfrac{P_L}{\sum_{i=1}^{N} I_{ti} - \dfrac{V_L{}^*}{R_L}}, & \forall i \in \{1,2,\dots,N\} \\ R_{eqi} = \dfrac{R_{pi}R_{qi}}{R_{pi}+R_{qi}} + R_{ti} \end{cases} \tag{37}$$

Then the stability of the model is determined by

$$\lambda(J(F(\dots I_{qi}\dots,\dots I_{ti}\dots,\dots V_{Ci}\dots,V_L))|_{V_L=V_L{}^*}) \tag{38}$$

where $\lambda(\cdot)$ represents eigenvalues and $J(\cdot)$ is the Jacobian matrix. If all eigenvalues have negative real parts, the system will be stable; otherwise, the system will be unstable.

In small-signal stability analysis, the stability region depends on the enormous parameters of the microgrid model, which may lead to the curse of high-dimensionality and the difficulty in visualization. Therefore, we study the influence of the parameters $P_L$ and $C_L$ on the stability region as an example to illustrate the difference between a small-signal stability region and a large-signal stability region. The small-signal stability region is shown in Fig. 14(a). The large-signal stability region is obtained using the novel proposed stability criteria, as depicted in Fig. 14(b).

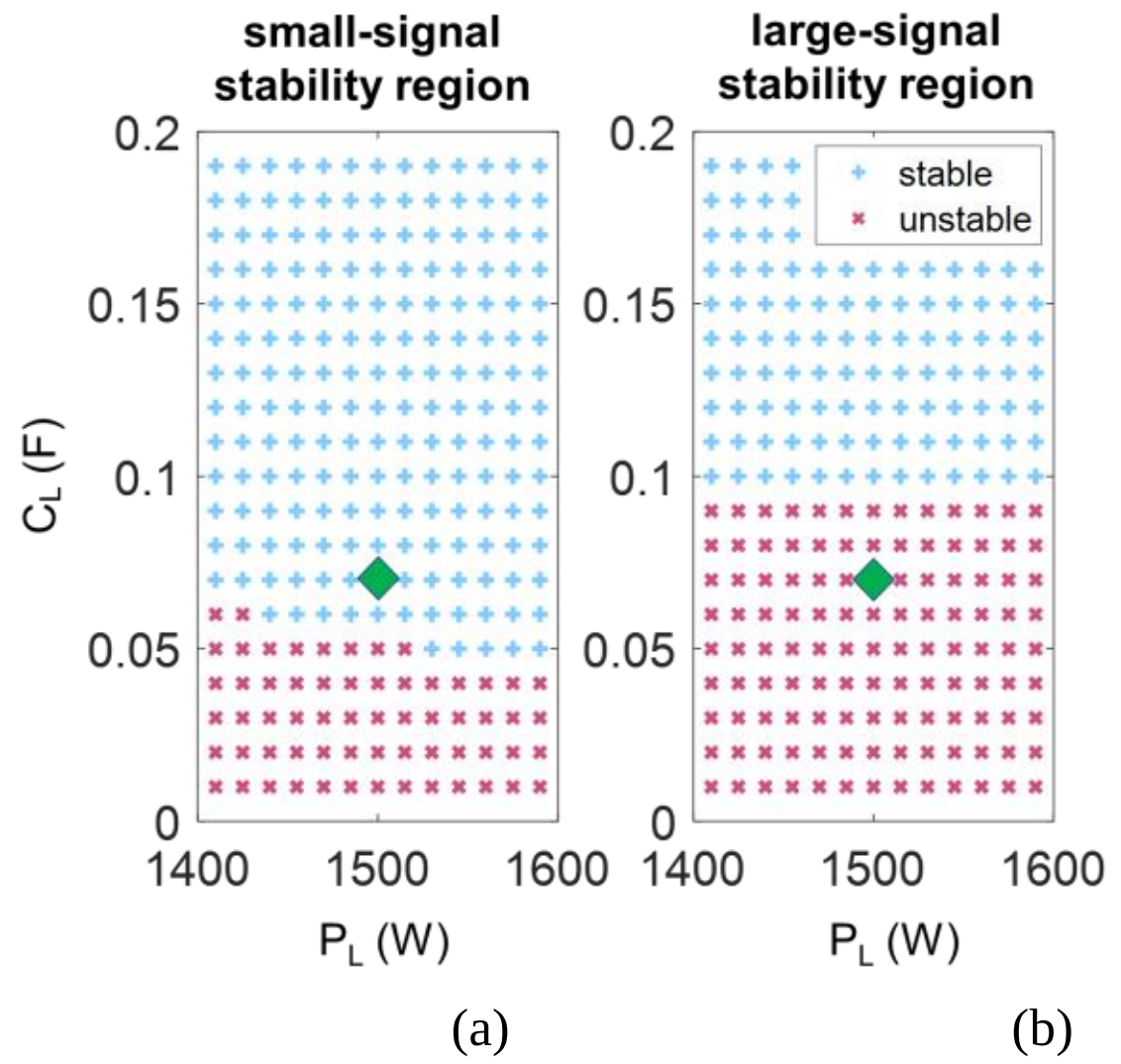

Fig. 14. (a) The small-signal stability region; (b) large-signal stability region.

A simulation is carried out at the data point $(P_L, C_L) = (1500W, 0.07F)$, which is marked as a green rhombus in Fig. 14. The dynamic response of the voltage at PoL is shown in Fig. 15. It is observed that the voltage oscillates severely after the CPL is plugged into the system. The simulation results show great correspondence with the theoretical results in Fig. 10 (b). We conclude that a small-signal stability region is not reliable in a DC microgrid; a large-signal stable system is naturally small-signal stable, but the opposite is hard to determine.

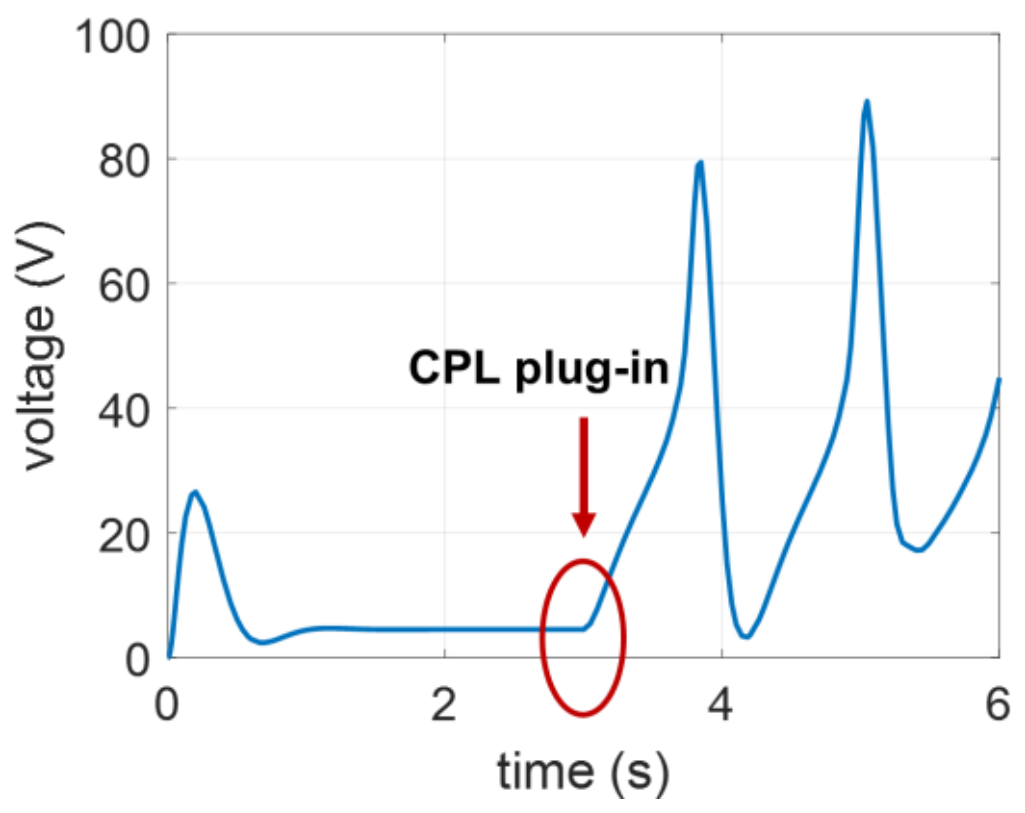

Fig. 15. The dynamic response of the voltage at PoL in a microgrid model.

## VI. Conclusion and Future Work

In this paper, we rigorously derive the sufficient criteria for large-signal stability in the DC microgrid with distributed-controlled power converters. To the best of the authors' knowledge, this systematic methodology is proposed for the first time. The acquisition of the sufficient criteria for global asymptotic stability is derived from Tellegen's theorem [22] and stability theories. Additionally, we present a novel distributed control method for power converters in a DC microgrid, which exhibits better performance than traditional droop control. The proposed controller is studied using its equivalent circuit model. Our future work will also investigate the performance comparison between our proposed controller and more advanced droop controllers such as [23]. Moreover, this paper reveals the defects of Brayton-Moser's mixed potential theory, which has been applied extensively since it was proposed in 1964. We also mention the important characteristics of the potential function, which are often utilized misleadingly in previous studies. Hardware tests will be implemented for further studies. Lastly, considering the fact that distributed generators can work in either current mode or voltage mode in practice, we will extend our research to fit for the DC microgrid with distributed generators in different operation modes and with more complicated interconnections.

## VII. Acknowledgment

The authors would like to thank the anonymous editors and reviewers for their valuable comments and suggestions to improve the quality of this paper.

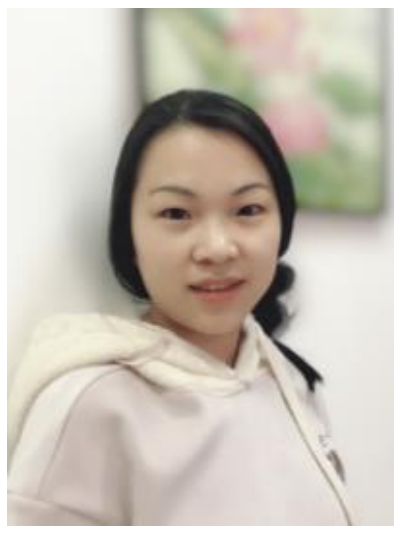

**Fangyuan Chang** (S'18) received the B.S. degree in electrical engineering from North China Electric Power University, China, in 2016, and the M.S. degree in electrical engineering from the University of Michigan, Ann Arbor, MI, in 2018. She is currently working toward the Ph.D. degree at the University of Michigan, Dearborn, MI. Her research interests include power systems, control theory, and machine learning.

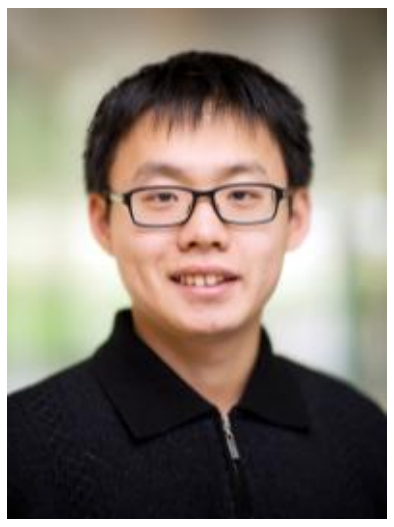

**Xiaofan Cui** (S'17) is currently a Ph.D. candidate in electrical engineering and computer science at the University of Michigan, Ann Arbor. He received the B.S. degree in electrical engineering and automation from Tsinghua University in 2016. He was a visiting student at Standard University during the summer of 2015. His research interests include circuit modeling, digital control, and high-performance power electronics.

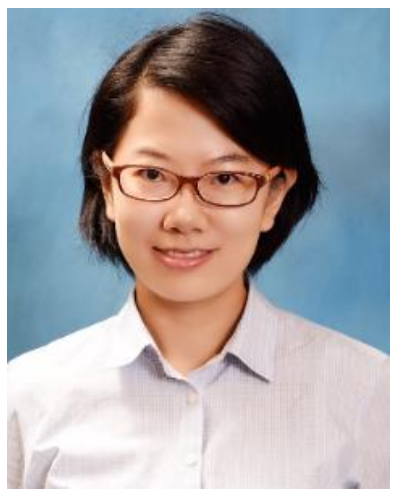

**Mengqi Wang** (S'11-M'15-SM'20) received the B.S. degree in electrical engineering from Xi'an Jiaotong University, Xi'an, in 2009 and the Ph.D. degree in electrical engineering from North Carolina State University, Raleigh, NC, in 2014. Since 2015, she has been an assistant professor with the Department of Electrical and Computer Engineering, University of Michigan-Dearborn, USA. She is a Senior Member of IEEE. Her current research interests include DC-DC and DC-AC power conversions, high efficiency and high power-density power supplies, renewable energy systems, and wide-bandgap power device applications.

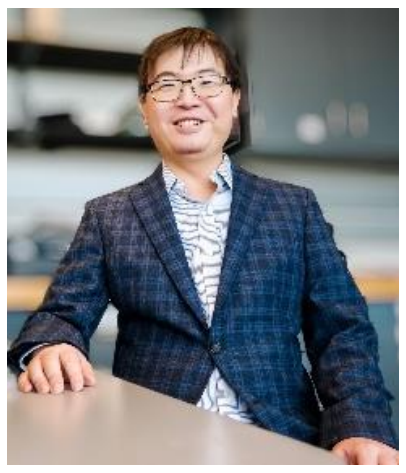

**Wengcong Su** (S'06-M'13-SM'18) received the B.S. degree (with distinction) from Clarkson University, Potsdam, NY, USA, in 2008, the M.S. degree from Virginia Tech, Blacksburg, VA, USA, in 2009, and the Ph.D. degree from North Carolina State University, Raleigh, NC, USA, in 2013, respectively. He is currently an Associate Professor in the Department of Electrical and Computer Engineering at the University of Michigan-Dearborn, USA. His current research interests include power systems, electrified transportation systems, and cyber-physical systems. He is a Fellow of IET and a Senior Member of IEEE. He is an Editor of IEEE Transactions on Smart Grid and an Associate Editor of IEEE Access. He is a registered Professional Engineer (P.E.) in the State of Michigan, USA. He has published more than 100 research papers in prestigious international journals and peer-reviewed conference proceedings. He is the recipient of 2015 IEEE Power and Energy Society (PES) Technical Committee Prize Paper Award and 2013 IEEE Industrial Electronics Society (IES) Student Best Paper Award.

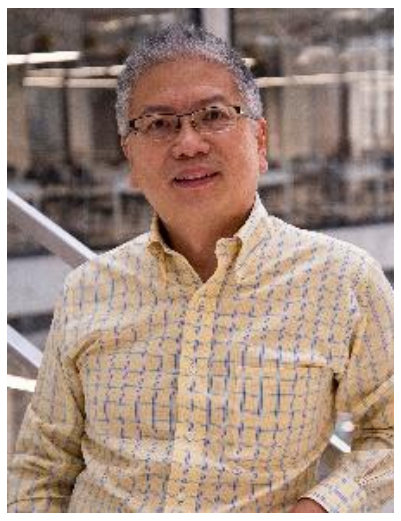

**Alex Q. Huang** (S'91-M'94-SM-96-F'05) was born in Zunyi, Guizhou, China. He received his B.Sc. degree from Zhejiang University, China in 1983, and his M.Sc. degree from Chengdu Institute of Radio Engineering, China in 1986, his Ph.D. from Cambridge University, the UK in 1992. From 1992 to 1994, he was a research fellow at Magdalene College, Cambridge. From 1994 to 2004, he was a professor at the Bradley Department of Electrical and Computer Engineering, Virginia Polytechnic Institute and State University, Blacksburg, Virginia. From 2004 to 2017, he was the Progress Energy Distinguished Professor of Electrical and Computer Engineering at NC State University where he established and led the NSF FREEDM Systems Center. Since 2017, he has become the Dula D. Cockrell Centennial Chair in Engineering at the University of Texas at Austin. Since 1983, he has been involved in the development of Si and WBG power semiconductor devices and power integrated circuits. He fabricated the first IGBT power device in China in 1985. He is the inventor and key developer of Si and SiC emitter turn-off (ETO) thyristor. He developed the concept of Energy Internet and the smart transformer-based Energy Router technology. He has mentored and graduated more than 90 Ph.D. and master students, and has published more than 550 papers in international conferences and journals. He has also been granted more than twenty U.S. patents. He is the recipient of the NSF CAREER award, the prestigious R&D 100 Award, the MIT Technology Review's 2011 Technology of the Year Award and the 2019 IEEE IAS Gerald Kliman Innovator Award. Dr. Huang is a fellow of the National Academy of Inventors (NAI) and IEEE.

## APPENDIX A.

The original dynamics of a complete circuit system is as follows:

$$-J\frac{dx}{dt}=\frac{\partial P(x)}{\partial x} \quad (1)$$

where $x=[I \quad V]^T, J=\begin{bmatrix}-L & 0\\ 0 & C\end{bmatrix}$.

Suppose a pair $(P^*,J^*)$ satisfies

$$J^*=\left(\lambda\mathbb{I}+\frac{\partial^2P(x)}{\partial x^2}M\right)\cdot J,\ P^*=\lambda P+\frac{1}{2}(\frac{\partial P(x)}{\partial x},M\frac{\partial P(x)}{\partial x}) \quad (2)$$

where $\mathbb{I}$ is an identity matrix, $M$ is a constant symmetric matrix, and $\lambda$ is a constant. Then we obtain:

$$\begin{aligned}-J^*\frac{dx}{dt}&=-\left(\lambda\mathbb{I}+\frac{\partial^2P(x)}{\partial x^2}M\right)\cdot J\frac{dx}{dt}\\&=-\lambda\cdot J\frac{dx}{dt}-\frac{\partial^2P(x)}{\partial x^2}M\cdot J\frac{dx}{dt}\end{aligned} \quad (3)$$

$$\begin{aligned}\frac{\partial P^*(x)}{\partial x}&=\frac{\partial}{\partial x}[\lambda\cdot P(x)+\frac{1}{2}(\frac{\partial P(x)}{\partial x},M\frac{\partial P(x)}{\partial x})]\\&=\lambda\cdot\frac{\partial P(x)}{\partial x}+\frac{\partial^2P(x)}{\partial x^2}M\frac{\partial P(x)}{\partial x}\\&=-\lambda\cdot J\frac{dx}{dt}-\frac{\partial^2P(x)}{\partial x^2}M\cdot J\frac{dx}{dt}\end{aligned} \quad (4)$$

Therefore, we can conclude

$$-J^*\frac{dx}{dt}=\frac{\partial P^*(x)}{\partial x} \quad (5)$$

where $J^*$ is always positive definite when the system is asymptotically stable.

## APPENDIX B.

**Theorem 2** Given a nonlinear circuit $\frac{dx}{dt}=f(x)$,

a) Let $P^*:\mathcal{R}^n\to R$ be of the class $C^1$ such that

i. $-J^*\frac{dx}{dt}=\frac{\partial P^*(x)}{\partial x}$ where $J^*\succ 0$

ii. $P^*(x)$ is radially unbounded, i.e., $P^*(x)\to\infty$ as $\|x\|\to\infty$

iii. $E:=\{x\in\mathcal{R}^n|f(x)=0\}$ , all equilibria of the nonlinear circuit are a compact set.

then every solution starting in $\mathcal{R}^n$ approaches $E$ as $t\to\infty$.

b) If $P^*$ is of class $C^2$ on the set $E$, let $M=\{\,x\in E|\,\frac{\partial^2P^*}{\partial x^2}\geq 0\}$, then every solution starting in $\mathcal{R}^n$ approaches $M$ as $t\to\infty$.

**Proof of Theorem 2a):**

Define: $c\triangleq\min_{x\in E}P^*(x)$. Since $E$ is a compact set, $c$ exists.
Since $P^*(x)$ is radially unbounded, given $c$, $\exists\gamma>0$, $s.t. P^*(x)>\ c$ where $\|x\|>\gamma$.
By contradiction, we know the $c$-level set of $P^*(x)$ $\Omega_c:=\{x\in\mathcal{R}^n|P^*(x)\leq\ c\}$ satisfies $\Omega_c\subset B_\gamma$ where $B_\gamma=\{x\in\mathcal{R}^n|\|x\|\leq\gamma\}$. Hence $\Omega_c$ is bounded.
Because $P^*(x)$ is defined in $\mathcal{R}^n$, by definition, we can easily see $\Omega_c$ is closed.
Because $\frac{dP^*(x)}{dt}\leq 0$, $\Omega_c$ is a compact and invariant set.
$E$ is the set of all points in $\Omega_c$ where $\frac{dP^*(x)}{dt}=0$.
From Lasalle's theorem [25], then every solution starting in $\Omega_c$ approaches $E$ as $t\to\infty$.
By increasing $c$ to infinity, we prove that every solution starting in $\mathcal{R}^n$ approaches $E$ as $t\to\infty$.

**Proof of Theorem 2b):**

We will prove that $M$ contains the largest invariant set in $E$, i.e., $\forall x_e\in E\backslash M,\ \frac{\partial^2P^*}{\partial x^2}\Big|_{x=x_e}\geq 0$ does not hold.
For simplicity, we denote $\frac{\partial^2P^*}{\partial x^2}\Big|_{x=x_e}$ by $H_{x_e}$.
Assume the eigenvalue decomposition of $H_{x_e}$: $H_{x_e}=U^T\Lambda U$, where $U$ is an orthogonal matrix and $\Lambda$ is a diagonal matrix.
There exists at least an entry $\lambda_i$ of $\Lambda$, $\lambda_i<0$. Without loss of generality, we consider $\lambda_1<0$.
Construct a function $V(x)$ as $V(x)=P^*(x_e)-P^*(x)$.
The Taylor expansion of $P^*(x)$ is

$$P^*(x)=P^*(x_e)+(\nabla P^*)^T|_{x=x_e}\hat{x}+\frac{1}{2}\hat{x}^TH_{x_e}\hat{x}+g(\hat{x}) \quad (6)$$

where $\hat{x}=x-x_e, g(\hat{x})=o(\|\hat{x}\|^2)$.

Substituting to $V(x)$:

$$V(\hat{x})=-\frac{1}{2}\hat{x}^TH_{x_e}\hat{x}-g(\hat{x}) \quad (7)$$

$$V(\hat{x})|_{\hat{x}=0}=0 \quad (8)$$

We select a set $\zeta_1=\{\hat{x}_p|\hat{x}_p=\mu U^Te_1, e_1=[1,0,0,\dots,0]^T,\mu\in\mathcal{R},\mu\neq 0\}$.

$$\begin{aligned}V(\hat{x})|_{\hat{x}\in\zeta_1}&=-\frac{1}{2}\mu^2({e_1}^T\Lambda e_1)-g(\hat{x}_p)\\&=-\frac{\lambda_1}{2}\ \|\hat{x}_p\|^2-g(\hat{x}_p)\end{aligned} \quad (9)$$

$\exists\, r_1, g(\hat{x})<-\frac{\lambda_1}{2}\ \|\hat{x}\|^2$ for all $\|\hat{x}\|^2\leq r_1$.

$$\dot{V}(\hat{x})|_{\hat{x}\in\zeta_1}=\dot{\hat{x}}J^*\dot{\hat{x}}\Big|_{\hat{x}\in\zeta_1}=\hat{x}^TH_x((J^*)^{-1})^TJ^*(J^*)^{-1}H_x\hat{x}|_{\hat{x}\in\zeta_1}$$
$$\Rightarrow\dot{V}(\hat{x})|_{\hat{x}\in\zeta_1}=\hat{x}^TH_x((J^*)^{-1})^TH_x\hat{x}|_{\hat{x}\in\zeta_1} \quad (10)$$

Denote $(H_x-H_{x_e})\hat{x}_p=\delta\hat{y}_p, H_{x_e}\hat{x}_p=\hat{y}_p$.

$$\dot{V}(\hat{x})|_{\hat{x}\in\zeta_1}=(\hat{y}_p+\delta\hat{y}_p)^T((J^*)^{-1})^T(\hat{y}_p+\delta\hat{y}_p) \quad (11)$$

**Lemma 1** Matrix $J^* \in \mathcal{R}^{n\times n}$ is positive definite (p.d.), then $(J^*)^T$ and $(J^*)^{-1}$ are also p.d.
Proof: $J^* \succ 0 \Leftrightarrow x^T J^* x > 0\ \forall x \neq 0 \Leftrightarrow (x^T J^* x)^T > 0\ \forall x \neq 0 \Leftrightarrow x^T (J^*)^T x > 0\ \forall x \neq 0 \Leftrightarrow (J^*)^T \succ 0$.
$$x^T (J^*)^{-1} x > 0\ \forall x \neq 0 \Leftrightarrow y^T (J^*)^T (J^*)^{-1} J^* y > 0\ \forall y \neq 0 \Leftrightarrow y^T (J^*)^T y > 0\ \forall y \neq 0$$

From Lemma 1, $(J^*)^{-1}$is positive definite.
$$\dot{V}(\hat{x})|_{\hat{x}\in\zeta_1} = (\hat{y}_p + \delta\hat{y}_p)^T K (\hat{y}_p + \delta\hat{y}_p) \quad (12)$$

where $K = \frac{1}{2}((J^*)^{-1} + ((J^*)^{-1})^T)$ is a p.d symmetric matrix.

$$\hat{y}_p = H_{x_e}\hat{x}_p = U^T \Lambda U(\mu U^T e_1) = \mu\lambda_1 U^T e_1 \quad (13)$$

$$\|\hat{y}_p\| = -\lambda_1 \|\hat{x}_p\| \quad (14)$$

$$\hat{y}_p{}^T K \hat{y}_p \geq \lambda_{min}(K)\|\hat{y}_p\|^2 > 0 \quad (15)$$

Since $H_x$ is continuous on $D$,
$$\exists r_2, \|H_x - H_{x_e}\|_M \leq -\lambda_1 \frac{\lambda_{min}(K)}{3\lambda_{max}(K)}, \forall\|x\| \leq r_2 \quad (16)$$

where $\|\cdot\|_M$ is the induced norm of the matrix.

$$\begin{aligned}\|\delta\hat{y}_p\| &= \|(H_x - H_{x_e})\hat{x}_p\| \leq \|H_x - H_{x_e}\|_M \|\hat{x}_p\| \\ &\leq -\lambda_1\|\hat{x}_p\| \frac{\lambda_{min}(K)}{3\lambda_{max}(K)} = \|\hat{y}_p\| \frac{\lambda_{min}(K)}{3\lambda_{max}(K)}\end{aligned} \quad (17)$$

$$\begin{aligned}&\delta\hat{y}_p{}^T K \hat{y}_p + \hat{y}_p K \delta\hat{y}_p{}^T + \delta\hat{y}_p{}^T K \delta\hat{y}_p{}^T \\ &\geq -\|\delta\hat{y}_p\|(2\|\hat{y}_p\| + \|\delta\hat{y}_p\|)\lambda_{max}(K) \\ &\geq -\frac{\lambda_{min}(K)}{3\lambda_{max}(K)}\|\hat{y}_p\|^2 \left(2 + \frac{\lambda_{min}(K)}{3\lambda_{max}(K)}\right)\lambda_{max}(K) \\ &> -\lambda_{min}(K)\|\hat{y}_p\|^2\end{aligned} \quad (18)$$

Therefore, we have
$$\dot{V}(\hat{x})|_{\hat{x}\in\zeta_1} > \lambda_{min}(K)\|\hat{y}_p\|^2 - \lambda_{min}(K)\|\hat{y}_p\|^2 = 0 \quad (19)$$

We define set $\mathcal{U} \triangleq \zeta_1 \cap B\left(\min(r_1, r_2)\right)$, where $B_\gamma$ represents $B_\gamma \triangleq \{\hat{x} \in \mathcal{R}^n | \|\hat{x}\| \leq \gamma\}$.

i. $V(\hat{x}) = 0$ at $\hat{x} = 0$
ii. $V(\hat{x}_p) > 0$ at some $\hat{x}_p = \mu U e_1$ with arbitrary small $\|\hat{x}_p\|$
iii. $\dot{V}(\hat{x}) > 0$ in $\mathcal{U}$

From Chetaev's theorem [26], $x = x_e$ is locally unstable.
The solution starting at $x(0) = x_e$ cannot stay identically in $E$; hence, $x_e$ is not included in the largest invariant set in $E$. Therefore, $M$ includes the largest invariant set in $E$.
From Lasalle's theorem, every solution starting in $\Omega_c$ approaches $M$ as $t \to \infty$.

By increasing $c$ to infinity, we prove that every solution starting in $\mathcal{R}^n$ approaches $M$ as $t \to \infty$.

## APPENDIX C.

The potential function of the system in Fig. 6 is

$$\begin{aligned}P(i,v) = &\sum_{i=1}^{N} V_{refi}\left(I_{pi} + I_{qi}\right) - \frac{1}{2}\sum_{i=1}^{N} R_{pi} I_{pi}{}^2 - \frac{1}{2}\sum_{i=1}^{N} R_{qi} I_{qi}{}^2 \\ &-\frac{1}{2}\sum_{i=1}^{N} R_{ti} I_{ti}{}^2 - \sum_{i=1}^{N} V_{Ci}\left(I_{pi} + I_{qi} - I_{ti}\right) - \frac{V_L{}^2}{2R_L} \\ &-V_L\left(\sum_{i=1}^{N} I_{ti} - \frac{P_L}{V_L} - \frac{V_L}{R_L}\right) + \int_{V_{min}}^{V_L} \frac{P_L}{v} dv - P_L\end{aligned} \quad (20)$$

where $V_L \geq V_{min}$. It can be simplified as follows:

$$\begin{aligned}P(i,v) = &\sum_{i=1}^{N} V_{refi}\left(I_{pi} + I_{qi}\right) - \frac{1}{2}\sum_{i=1}^{N} R_{pi} I_{pi}{}^2 - \frac{1}{2}\sum_{i=1}^{N} R_{qi} I_{qi}{}^2 \\ &-\frac{1}{2}\sum_{i=1}^{N} R_{ti} I_{ti}{}^2 - \sum_{i=1}^{N} V_{Ci}\left(I_{pi} + I_{qi} - I_{ti}\right) \\ &+\frac{V_L{}^2}{2R_L} - V_L \sum_{i=1}^{N} I_{ti} + \int_{V_{min}}^{V_L} \frac{P_L}{v} dv\end{aligned} \quad (21)$$

Define the following notations:

$R = diag([R_{p1}, \dots, R_{pN}, R_{q1}, \dots, R_{qN}, R_{t1}, \dots, R_{tN}])$
$= diag([R_p, R_q, R_t])$,
$L = diag([L_{p1}, \dots, L_{pN}, L_{q1}, \dots, L_{qN}, L_{t1}, \dots, L_{tN}])$
$= diag([L_p, L_q, L_t])$,
$C = diag([C_{b1}, C_{b2}, \dots, C_{bN}, C_L]) = diag([C_b, C_L])$,
$i = [I_{p1}, \dots, I_{pN}, I_{q1}, \dots, I_{qN}, I_{t1}, \dots, I_{tN}]_{3N\times 1}$,
$v = [V_{C1}, V_{C2}, \dots, V_{CN}, V_L]_{(N+1)\times 1}$.

Then we rewrite the potential function in equation (21) in the form of
$$P(i,v) = -A(i) + B(v) + (i, \gamma v - a), \quad (22)$$

where $A: \mathbb{R}^{3N} \to \mathbb{R}, B: \mathbb{R}^{N+1} \to \mathbb{R}, \gamma$ is a constant matrix and $a$ is a constant vector, $(\cdot,\ \cdot)$ represents an inner product. Then we obtain that

$$A = \frac{1}{2} i^T R i, \qquad \gamma = \begin{bmatrix} -\mathbb{I}_{N\times N} & 0_{N\times 1} \\ -\mathbb{I}_{N\times N} & 0_{N\times 1} \\ \mathbb{I}_{N\times N} & -1_{N\times 1} \end{bmatrix}_{(3N)\times(N+1)} \quad (23)$$

where $\mathbb{I}$ is an identity matrix. Therefore, we have

$$L^{\frac{1}{2}}A^{-1}\gamma C^{-\frac{1}{2}} = \begin{bmatrix} -L_p^{\frac{1}{2}}R_p^{-1}C_b^{-\frac{1}{2}} & 0_{N\times 1} \\ -L_q^{\frac{1}{2}}R_q^{-1}C_b^{-\frac{1}{2}} & 0_{N\times 1} \\ -L_t^{\frac{1}{2}}R_t^{-1}C_b^{-\frac{1}{2}} & -L_t^{\frac{1}{2}}R_t^{-1}C_L^{-\frac{1}{2}} \end{bmatrix}_{3N\times(N+1)} \quad (24)$$

Specifically, considering the virtual inductor $L_p = 0$, $L^{1/2}A^{-1}\gamma C^{-1/2}$ can be simplified as

$$L^{\frac{1}{2}}A^{-1}\gamma C^{-\frac{1}{2}} = \begin{bmatrix} 0_{N\times N} & 0_{N\times 1} \\ -L_q^{\frac{1}{2}}R_q^{-1}C_b^{-\frac{1}{2}} & 0_{N\times 1} \\ -L_t^{\frac{1}{2}}R_t^{-1}C_b^{-\frac{1}{2}} & -L_t^{\frac{1}{2}}R_t^{-1}C_L^{-\frac{1}{2}} \end{bmatrix}_{3N\times(N+1)} \quad (25)$$

One condition for global stability from the Theorem 3 in [14] is

$$\|L^{1/2}A^{-1}\gamma C^{-1/2}\| \le 1-\delta, \delta > 0 \quad (26)$$

Considering that $\|L^{1/2}A^{-1}\gamma C^{-1/2}\|$ can be solved by the largest singular value of $L^{1/2}A^{-1}\gamma C^{-1/2}$, we obtain the first condition for large-signal stability as follows:

$$\sigma_{max}\left(\begin{bmatrix} 0_{N\times N} & 0_{N\times 1} \\ -L_q^{\frac{1}{2}}R_q^{-1}C_b^{-\frac{1}{2}} & 0_{N\times 1} \\ -L_t^{\frac{1}{2}}R_t^{-1}C_b^{-\frac{1}{2}} & -L_t^{\frac{1}{2}}R_t^{-1}C_L^{-\frac{1}{2}} \end{bmatrix}\right) < 1 \quad (27)$$

where $\sigma_{max}(\cdot)$ is the largest singular value.

## APPENDIX D.

**Part I. The stability region derived from Brayton-Moser's mixed potential theory**

The applied theorem from [14] is introduced first.

**Theorem** Consider the potential of a dynamic system

$$P(i,v) = -\frac{1}{2}(i,Ai) + B(v) + (i,\gamma v - \alpha) \quad (28)$$

If $A$ is positive definite, $B(v) + |\gamma v| \to \infty$ as $|v| \to \infty$, and

$$\|L^{1/2}A^{-1}\gamma C^{-1/2}\| \le 1-\delta, \delta > 0 \quad (29)$$

for all $i, v$, then all solutions of the system $-J\frac{dx}{dt} = \frac{\partial P(x)}{\partial x}$ tend to the set of equilibrium points as $t \to \infty$.

**Solution:**
The potential function of the circuit model in Fig. 13 is

$$P(I_L,V_C) = V_sI_L - \frac{1}{2}RI_L^2 - V_C\left(I_L - \frac{V_C}{R_L}\right) - \frac{V_C^2}{2R_L}$$
$$= V_sI_L - \frac{1}{2}RI_L^2 - V_CI_L + \frac{V_C^2}{2R_L} \quad (30)$$

Rewrite $P(I_L, V_C)$ in the following form:

$$P(i,v) = -\frac{1}{2}(i,Ai) + B(v) + (i,\gamma v - \alpha) \quad (31)$$

where $i = I_L, v = V_C, A = R, B(v) = \frac{V_C^2}{2R_L}, \gamma = -1, \alpha = -V_s$.

We know $A = R \succ 0$. Moreover, we have

$$\|L^{1/2}A^{-1}\gamma C^{-1/2}\| < 1 \Rightarrow \frac{1}{R}\sqrt{\frac{L}{C}} < 1 \quad (32)$$

$$B(v) + |\gamma v| = \frac{V_C^2}{2R_L} + V_C \quad (33)$$

However, because $R_L < 0$, $B(v) + |\gamma v| \nrightarrow \infty$ as $|V_C| \to \infty$. Therefore, the obtained stability region is $\{\phi\}$, i.e. an empty set.

**Part II. The stability region derived from our proposed criteria**

**Solution:**
The potential function $P(I_L, V_C)$ of the circuit shown in Fig.13 is:

$$P(I_L,V_C) = V_sI_L - \frac{1}{2}RI_L^2 - V_CI_L + \frac{V_C^2}{2R_L} \quad (34)$$

Rewrite $P(I_L, V_C)$ in the following form:

$$P(i,v) = -\frac{1}{2}(i,Ai) + B(v) + (i,\gamma v - \alpha) \quad (35)$$

where $i = I_L, v = V_C, A = R, B(v) = \frac{V_C^2}{2R_L}, \gamma = -1, \alpha = -V_s$.
Review the proposed stability criteria in section IV:
a. $P^*(x)$ is radially unbounded, i.e., $P^*(x) \to \infty$ as $\|x\| \to \infty$.
b.

$$\begin{cases} \sigma_{max}(L^{1/2}A^{-1}\gamma C^{-1/2}) < 1 \\ \left.\frac{\partial^2 B(v)}{\partial v^2} + \gamma^T A^{-1}\gamma\right|_{v=v_e} \ge 0 \end{cases} \quad (36)$$

where $\sigma_{max}(\cdot)$ is the largest singular value.
In the circuit shown in Fig. 13,

$$\sigma_{max}(L^{1/2}A^{-1}\gamma C^{-1/2}) < 1 \Rightarrow \frac{1}{R}\sqrt{\frac{L}{C}} < 1 \quad (37)$$

$$\left.\frac{\partial^2 B(v)}{\partial v^2} + \gamma^T A^{-1}\gamma\right|_{v=v_e} \ge 0 \Rightarrow \frac{1}{R_L} + \frac{1}{R} \ge 0 \Rightarrow R \le |R_L| \quad (38)$$

According to equation (38), equation (37) can be converted to

$$\frac{L}{C} < R \cdot |R_L| \quad (39)$$

From equations (38)(39), we have

$$\frac{L}{|R_L|C} < R \le |R_L| \quad (40)$$

Next, it remains to be proved that $P^*(i,v) \to \infty$ as $|i| + |v| \to \infty$. The potential function is

$$P(I_L, V_C) = V_s I_L - \frac{1}{2} R I_L{}^2 - V_C I_L + \frac{V_C{}^2}{2R_L} \tag{41}$$

Choose $\lambda = 1$, $M = \begin{bmatrix} 2A^{-1} & 0 \\ 0 & 0 \end{bmatrix}$. Notate $\frac{\partial P}{\partial V_C}$, $\frac{\partial P}{\partial I_L}$ and $\frac{\partial^2 B(v)}{\partial v^2}$ by $P_{V_C}$, $P_{I_L}$, and $B_{vv}(v)$, separately. Suppose $\mu_1$ is the smallest eigenvalue of the matrix $L^{-1/2}A\,(i)L^{-1/2}$ for all $i$, and $\mu_2$ is the smallest eigenvalue of the matrix $C^{-1/2}B_{vv}(v)C^{-1/2}$ for all $v$. Then we have:

$$P^*(I_L, V_C) = \left(\frac{\mu_1 - \mu_2}{2}\right) P(I_L, V_C) + \frac{1}{2}\left(P_{I_L}, L^{-1} P_{I_L}\right) + \frac{1}{2}\left(P_{V_C}, C^{-1} P_{V_C}\right) \tag{42}$$

where

$$\mu_1 = \min\left\{\lambda\left(L^{-\frac{1}{2}} A\,(i) L^{-\frac{1}{2}}\right)\right\} = \frac{R}{L} \tag{43}$$

$$\mu_2 = \min\left\{\lambda\left(C^{-\frac{1}{2}} B_{vv}(v) C^{-\frac{1}{2}}\right)\right\} = \frac{1}{CR_L} \tag{44}$$

Plugging in the value of $\mu_1$ and $\mu_2$, we have

$$P^*(I_L, V_C) = \left(\frac{R}{L} - \frac{1}{CR_L}\right)\left(V_s I_L - \frac{1}{2} R I_L{}^2 - V_C I_L + \frac{V_C{}^2}{2R_L}\right) + \frac{1}{2L}(RI_L + V_C - V_s)^2 + \frac{1}{2C}\left(\frac{V_C}{R_L} - I_L\right)^2 \tag{45}$$

Suppose $P^* = \frac{1}{2} x^T P_2 x + P_1{}^T x + P_0$, where

$$P_2 = \begin{bmatrix} R & 1 \\ 1 & \frac{2}{R} + \frac{1}{R_L} \end{bmatrix}, P_1 = \begin{bmatrix} -V_s \\ -\frac{2}{R} V_s \end{bmatrix}, P_0 = \frac{V_s{}^2}{2L}, x = [I_L \quad V_C]^T.$$

Denote the smallest eigenvalue of $P_2$ by λ. Since $\frac{\partial^2 P^*(x)}{\partial x^2} = P_2 \geq 0$, we have $\lambda \geq 0$.

It is proved in the Courant–Fischer–Weyl min-max principle that

$$\frac{(Ax, x)}{(x, x)} \geq \lambda_{min} \tag{46}$$

where $A$ is a $n \times n$ symmetric matrix, $\lambda_{min}$ is the smallest eigenvalue of $A$.

According to the Courant–Fischer–Weyl min-max principle, we have

$$2(P^* - P_1{}^T x - P_0) \geq \lambda\left(I_L{}^2 + V_C{}^2\right) \tag{47}$$

Then

$$P^* \geq P_1{}^T x + P_0 + \frac{\lambda}{2}\left(I_L{}^2 + V_C{}^2\right) = \frac{\lambda}{2}\left(I_L{}^2 + V_C{}^2\right) - V_s(I_L + \frac{2}{R} V_C) + \frac{V_s{}^2}{2L} \tag{48}$$

The Cauchy–Schwarz inequality states that for all vectors $u$ and $v$ of an inner product space it is true that

$$\|u\| \cdot \|v\| \geq |(u, v)|, \tag{49}$$

where $\|\cdot\|$ is the norm of a vector.

Using Cauchy–Schwarz inequality, we have

$$\left\|\begin{bmatrix} 1 \\ 2/R \end{bmatrix}\right\| \cdot \left\|\begin{bmatrix} I_L \\ V_C \end{bmatrix}\right\| \geq \left|\left(\begin{bmatrix} 1 \\ 2/R \end{bmatrix}, \begin{bmatrix} I_L \\ V_C \end{bmatrix}\right)\right|, \tag{50}$$

i.e.,

$$\sqrt{1 + \left(\frac{2}{R}\right)^2} \cdot \sqrt{\left(I_L{}^2 + V_C{}^2\right)} \geq I_L + \frac{2}{R} V_C, \tag{51}$$

Since

$$(|I_L| + |V_C|)^2 = I_L{}^2 + V_C{}^2 + 2|I_L| \cdot |V_C| \geq I_L{}^2 + V_C{}^2, \tag{52}$$

we have

$$|I_L| + |V_C| \geq \sqrt{\left(I_L{}^2 + V_C{}^2\right)}. \tag{53}$$

Therefore,

$$\sqrt{1 + \left(\frac{2}{R}\right)^2} \cdot (|I_L| + |V_C|) \geq \sqrt{1 + \left(\frac{2}{R}\right)^2} \cdot \sqrt{\left(I_L{}^2 + V_C{}^2\right)} \geq \left(I_L + \frac{2}{R} V_C\right). \tag{54}$$

Therefore (48) can be converted to

$$P^* \geq \frac{\lambda}{2}\left(I_L{}^2 + V_C{}^2\right) - V_s\left(I_L + \frac{2}{R} V_C\right) + \frac{V_s{}^2}{2L} \geq \frac{\lambda}{4}(|I_L| + |V_C|)^2 - V_s \sqrt{1 + \left(\frac{2}{R}\right)^2} \cdot (|I_L| + |V_C|) + \frac{V_s{}^2}{2L} = (|I_L| + |V_C|) \cdot [\frac{\lambda}{4}(|I_L| + |V_C|) - V_s \sqrt{1 + \left(\frac{2}{R}\right)^2}] + \frac{V_s{}^2}{2L} \tag{55}$$

**If λ>0:** when $|I_L| + |V_C| \to \infty$, it is concluded that $P^* \to \infty$.

**If λ=0:** when $|I_L| + |V_C| \to \infty$, we cannot conclude $P^* \to \infty$.

Therefore, we need to rule out the case that $\lambda = 0$ and guarantee $\lambda > 0$, where λ is the smallest eigenvalue of $P_2$.

Let $\lambda > 0$ we obtain

$$R < -R_L \tag{56}$$

Combining equations (40)(56) we have:

$$\frac{L}{|R_L|C} < R < |R_L| \tag{57}$$

In conclusion, the stability region derived from our proposed stability criteria is as follows:

$$\frac{L}{|R_L|C} < R < |R_L| \tag{58}$$

QED.